\documentstyle[11pt]{article}

\if@twoside\oddsidemargin25mm
\else\oddsidemargin25mm\evensidemargin25mm\marginparwidth25mm\fi
\def\thefootnote{\dag}

\def\bl{\Big\{}                  \def\br{\Big\}}
\def\bpl{\Big(}                  \def\bpr{\Big)}
\def\bq{\begin{equation}}        \def\eq{\end{equation}}
\def\brr{\begin{eqnarray}}       \def\err{\end{eqnarray}}
\def\ba{\left(\begin{array}}     \def\ea{\end{array}\right)}
\def\der{\partial}
\def\Dd{{\cal D}}                \def\Ll{{\cal L}}
\def\Vv{{\cal V}}                \def\Oo{{\cal O}}
\def\Uu{{\cal U}}                \def\Aa{{\cal A}}
\def\Bb{{\cal B}}                \def\Cc{{\cal C}}

\def\g{\gamma}                   \def\d{\delta}
\def\e{\epsilon}                 \def\s{\sigma}
\def\t{\theta}                   \def\O{\Omega}
\def\ve{\varepsilon}             \def\l{\lambda}
\def\a{\alpha}                   \def\b{\beta}
\def\vphi{\displaystyle \varphi}

\def\bX{{\bar X}}                \def\bO{{\bar\Omega}}
\def\be{\bar\epsilon}            
\def\bpsi{\bar\psi}              \def\bF{\bar F}
\def\bZ{{\bar Z}}

\def\hF{{\hat F}}                \def\hG{{\hat G}}
\def\hC{{\hat C}}                \def\hH{{\hat H}}
\def\cF{{\check F}}              \def\cG{{\check G}}
              
\def\cW{{\check W}}              \def\ct{{\check\theta}}
\def\cX{{\check X}}              \def\cO{{\check\Omega}}
\def\cL{{\check\Lambda}}         \def\cY{{\check Y}}
\def\cB{{\check B}}              
\def\cbX{{\,\check{\!\bar X}}}   \def\cbO{{\check{\bar \O}}}
\def\hcF{{\,\hat{\!\check F}}}   \def\cZ{{\check Z}}
\def\cbZ{{\,\check{\!\bar Z}}}
\def\dslash{\hbox{\ooalign{$\displaystyle\partial$\cr$/$}}}
\def\Dslash{\hbox{\ooalign{$\displaystyle D$\cr$\hspace{.03in}/$}}}
\def\Vslash{\hbox{\ooalign{$\displaystyle V$\cr$\hspace{.02in}/$}}}
\def\hHslash{\hbox{\ooalign{$\displaystyle \hH$\cr$\hspace{.02in}/$}}}
\def\Vvslash{\hbox{\ooalign{$\displaystyle \Vv$\cr$\hspace{.02in}/$}}}

\newcommand{\IP}{\relax{\rm I\kern-.18em P}}
\newcommand{\dr}{\raise.3ex\hbox{$\stackrel{\leftarrow}{\partial }$}{}}
\newcommand{\dl}{\raise.3ex\hbox{$\stackrel{\rightarrow}{\partial}$}{}}
\newcommand{\eqn}[1]{(\ref{#1})}
\newcommand{\ft}[2]{\mbox{\Large $\frac{#1}{#2}$}}
\newcommand{\ef}[1]{e_{\phantom i}^{#1 \vphi} }
\newcommand{\eft}[1]{e^{#1 \mbox{\Large $\frac{\varphi}{2}$}}}

\newsavebox{\uuunit}
\sbox{\uuunit}
    {\setlength{\unitlength}{0.825em}
     \begin{picture}(0.6,0.7)
        \thinlines
        \put(0,0){\line(1,0){0.5}}
        \put(0.15,0){\line(0,1){0.7}}
        \put(0.35,0){\line(0,1){0.8}}
       \multiput(0.3,0.8)(-0.04,-0.02){12}{\rule{0.5pt}{0.5pt}}
     \end {picture}}
\newcommand {\unity}{\mathord{\!\usebox{\uuunit}}}
\newcommand {\Rbar} {{\mbox{\rm$\mbox{I}\!\mbox{R}$}}}

\newsavebox{\zzzbar}
\sbox{\zzzbar}
  {\setlength{\unitlength}{0.9em}
  \begin{picture}(0.6,0.7)
  \thinlines
  \put(0,0){\line(1,0){0.6}}
  \put(0,0.75){\line(1,0){0.575}}
  \multiput(0,0)(0.0125,0.025){30}{\rule{0.3pt}{0.3pt}}
  \multiput(0.2,0)(0.0125,0.025){30}{\rule{0.3pt}{0.3pt}}
  \put(0,0.75){\line(0,-1){0.15}}
  \put(0.015,0.75){\line(0,-1){0.1}}
  \put(0.03,0.75){\line(0,-1){0.075}}
  \put(0.045,0.75){\line(0,-1){0.05}}
  \put(0.05,0.75){\line(0,-1){0.025}}
  \put(0.6,0){\line(0,1){0.15}}
  \put(0.585,0){\line(0,1){0.1}}
  \put(0.57,0){\line(0,1){0.075}}
  \put(0.555,0){\line(0,1){0.05}}
  \put(0.55,0){\line(0,1){0.025}}
  \end{picture}}
\newcommand{\Zbar}{\mathord{\!{\usebox{\zzzbar}}}}
\begin{document}

\begin{titlepage}
\begin{flushright}  ULB-TH-97/17 \\ hep-th/9709129
\end{flushright}
\vfill
\begin{center}
{\LARGE\bf The low-energy effective action}
\vspace{4mm}

{\LARGE\bf for perturbative heterotic strings on $K_3 \times T^2$}
\vspace{4mm}

{\LARGE\bf and the d=4 N=2 vector-tensor multiplet}
\vfill
{\large  R. Siebelink}\\
\vspace{7mm}
{\small
     Service de Physique Th\'{e}orique, Universit\'{e} Libre de
     Bruxelles,\\
     Campus Plaine, CP 225, Bd du Triomphe,\\B--1050 Bruxelles, Belgium}
\end{center}
\vfill
\begin{center} {\bf Abstract}\end{center}
\vfill
{\small
We consider $d=4$ $N=2$ supergravity theories which serve as low-energy
effective actions for heterotic strings on $K_3 \times T^2$. At the
perturbative level we construct a new version of the heterotic effective
action in which the axion has been traded for an antisymmetric tensor
field. In the string frame the antisymmetric tensor doesn't transform 
under Poincar\'e supersymmetry into the dilaton-dilatini system. This 
indicates that in this frame the antisymmetric tensor field and the dilaton
are not contained in an $N=2$ vector-tensor multiplet. Instead, we find 
that the heterotic dilaton is part of a compensating hypermultiplet, 
whereas the antisymmetric tensor is part of the gravitational multiplet. 
In order to obtain our results we use superconformal techniques. This 
enables us to comment on the range of applicability of this particular 
framework.}
\vfill
\flushleft{}
\end{titlepage}
\def\thefootnote{\dag}
\renewcommand\thefootnote{\arabic{footnote}}
\section{Introduction}

During the last few years much progress has been made in the study of 
supersymmetric field
theories and string theories in various dimensions thanks to the discovery
of a whole set of duality symmetries. Four dimensional models with $N=2$
supersymmetry have proven to be particularly interesting because of the
rich physical phenomena that appear in this context. Moreover these
phenomena can be studied in great detail thanks to the specific structure
imposed by the $N=2$ supersymmetry. A by-now famous example concerns the
duality that relates heterotic strings on $K_3\times T^2$ to type IIA
strings on $K_3$-fibered Calabi-Yau manifolds \cite{KV,FHSV,KLM,AL}.
This duality has very powerful consequences. Suppose for instance that one
is describing the low-energy dynamics of these strings in terms
of an effective $N=2$ supergravity action with vector and hypermultiplet
couplings. The vector multiplet sector of such an effective action can in
principle be determined exactly by performing a tree level computation in
the type II picture. Thanks to the heterotic-type II duality one can 
reinterprete this exact result in a heterotic context. This automatically
solves a strong coupling problem because in terms of the heterotic
variables the vector multiplet action receives tree level, one-loop and
non-perturbative contributions.

The interpretation of the type II result in a heterotic language is not so
obvious however, because the variables that are naturally inherited from
the type II side turn out to be complicated functions of the ``natural"
heterotic variables. The latter transform in a simple way under the
$SO(2,n)$ T-duality group, whereas the type II variables in general don't.
Amongst other things this implies that one needs to perform a whole set of
field redefinitions before the effective action can really be compared to 
direct perturbative heterotic string computations. Consider for instance 
the dilaton-axion-like scalar field $S= \phi - i a$ which on the type II 
side is defined as the complexified K\"ahler modulus of the $\IP^1$ base of
the $K_3$-fibration. This field $S$ is an $N=2$ ``special coordinate" and 
it is shifted by a purely imaginary constant under a (quantised) 
Peccei-Quinn symmetry. It is well-known \cite{dWKLL,BHW} that $S$ 
is not invariant under $SO(2,n)$ transformations once loop and 
non-perturbative contributions are taken into account. This 
indicates that $\phi$, which coincides with the true heterotic dilaton at 
the string tree level, starts to differ from it at the one-loop level or 
non-perturbatively. Therefore it is necessary to express the field $\phi$ 
as a function of the true dilaton $\phi_{\rm inv}$ and the other moduli, 
before one can properly separate the non-perturbative effects from the 
perturbative ones.  It is also known \cite{CDFVP} that
one has to perform a change of variables at the level of the vector fields.
The reason is that the $SO(2,n)$ transformations mix the field strengths 
for the type II inherited vectors\footnote{Incidently we will refer to 
these vector fields as the $STU$ vectors, because they lead to a vector 
multiplet action which is characterised by a prepotential $F(X)$ of the 
form $STU$ + more. Here $T$ and $U$ stand for the moduli of the heterotic 
$T^2$.} with their duals, which implies that these vectors themselves 
transform in a non-local way. In order to avoid this, one can perform an 
electro-magnetic duality transformation on one of the vectors, such that one
ends up with a new set of ``stringy" vectors which transform just linearly
into eachother under $SO(2,n)$.

In addition to the above-mentioned field redefinitions there is a last
change of variables which so far has been less well under control, and
which we intend to study in the course of the present work. What
we have in mind here is the duality transformation which trades the axion
$a$ for an antisymmetric tensor field $\cB_{\mu\nu}$. Of course this 
transformation can only be implemented after going to the perturbative region
of the vector multiplet moduli space, where the Peccei-Quinn symmetry is
continuous instead of being quantised. The axion can then be identified
with the zero form gauge potential associated to this continuous symmetry
and as such it can be dualised into an antisymmetric tensor field. In
\cite{dWKLL} it was conjectured that in the $N=2$ context this duality
transformation would replace the vector multiplet which originally
contained the scalar $S$ by a so-called {\it vector-tensor} multiplet 
\cite{vtrefs,dWKLL} in which $\phi_{\rm inv}$ and $\cB_{\mu\nu}$ would 
find their natural place.
In order to test this conjecture a systematic study of vector-tensor 
supergravities was undertaken in \cite{wij}\footnote{More recently 
superspace descriptions of the vector-tensor multiplet have appeared in 
\cite{vtrefs2}.}. An interesting class of 
interacting vector-tensor theories came out of this analysis, but quite 
surprisingly the sought-after heterotic vector-tensor theory was not found.
In this article we further investigate the issue of a possible 
vector-tensor structure in the antisymmetric tensor effective action for 
heterotic strings. We do this by explicitly constructing this antisymmetric
tensor effective action, together with its associated $N=2$ supersymmetry 
transformation rules. The outcome is first of all that we have to be 
careful before drawing rigorous conclusions about the possible 
(non)existence of the heterotic vector-tensor multiplet because the $N=2$ 
supersymmetry of our final model is only realised on-shell. This means 
that one first has to decide on which variables one uses as the 
fundamental ones, before one can identify the kind of multiplets these 
variables belong to. When the string metric and the corresponding string 
gravitinos are used as fundamental variables, the 
antisymmetric tensor $\cB_{\mu\nu}$ is {\it not} linked to dilaton-dilatini
system by supersymmetry. As a result there is no vector-tensor multiplet in
the string frame theory. 

In this article we also wish to comment on other issues which are
of interest from a purely supergravity point of view. Since it is not
always possible (nor desirable) to keep the heterotic string and 
supergravity ideas completely separated throughout the main text, we 
summarise the different lines of thought already here, for the sake of 
clarity.

\subsection{Construction of the antisymmetric tensor effective action for
perturbative heterotic strings with N=2 supersymmetry}

Our strategy for obtaining the antisymmetric tensor effective action 
for heterotic strings consists of performing a sequence of duality 
transformations. As a starting point we take a conventional $N=2$ 
supergravity theory coupled to a set of vector and hypermultiplets, and we
use the well-known vector multiplet prepotential for type II strings on 
$K_3$-fibered Calabi-Yau manifolds. This yields what we call the $STU$
version of the heterotic effective action. We review the basic properties 
of this model including its Peccei-Quinn and $SO(2,n)$ symmetries.
We briefly discuss how the $SO(2,n)$ symmetry can be made 
manifest at the lagrangian level by implementing the higher-mentioned 
duality transformation on the vector gauge fields \cite{CDFVP}. 
 
Before we dualise the axion we broaden our point of view and 
study generic vector multiplet theories containing a Peccei-Quinn symmetry.
We show that the class of Peccei-Quinn invariant models is in fact quite
restricted and precisely comprises the cases discussed in \cite{wij} plus
the case which is relevant for perturbative heterotic strings. The various
Peccei-Quinn invariant models can be dualised in a unified way and this 
explains why the resulting antisymmetric tensor theories share a similar 
gauge structure. Most notably there exists a particular $U(1)$ 
gauge symmetry which acts as a shift symmetry on the antisymmetric tensor.
For those cases that allow for an off-shell treatment along the lines of 
\cite{wij} this $U(1)$ transformation is nothing but the central charge 
transformation. We show that in the cases 
of \cite{wij} the central charge-like structure is really indispensable 
(even on-shell!) whereas this is not so for the heterotic antisymmetric 
tensor theory. In the latter case the central charge-like structure can be 
completely removed, and one ends up with a set of vector fields which all
appear on an equal footing. This of course reflects the underlying 
$SO(2,n)$ invariance of the heterotic model. 

It is worth mentioning that very little string information is used in our 
construction of the ``heterotic" antisymmetric tensor theory. In fact we 
just start from the most general Peccei-Quinn invariant vector multiplet 
theory and then select the case which contains an $SO(2,n)$ symmetry. At 
the end of the whole dualisation procedure we find that several known 
string theoretical properties are manifestly realised in our model, which 
shows that in fact they can be thought off as being a consequence of $N=2$
supersymmetry alone, rather then being of an intrinsic stringy nature. First
of all we nicely reproduce the specific couplings of the dilaton and the 
antisymmetric tensor to the other moduli. Furthermore we see that the 
supersymmetry transformation rules for all the fields are (almost) 
completely independent of the one-loop part of the theory. In this respect
the antisymmetric tensor formulation is simpler then the other possible 
formulations of the heterotic effective action in terms of vector and 
hypermultiplets only.

\subsection{N=2 superconformal supergravity and the perturbative
heterotic string}

It so happens that the low-energy effective action for perturbative
heterotic strings is a perfect laboratory to address some interesting
supergravity issues. These supergravity issues form a second main
ingredient of this paper. In particular  we focus on the superconformal
framework for $d=4$ $N=2$ supergravity \cite{tensorcal,dWVP,dWLVP} which is
very well suited for our present purposes. 

The general philosophy of the superconformal approach to $N=2$ supergravity
is the following. Although the ultimate goal of the superconformal 
techniques is to construct super Poincar\'e theories describing the 
on-shell interactions of a certain set of physical fields, one starts off 
with various multiplets that form a representation of the {\it off-shell 
superconformal algebra} (which is considerably larger then the super 
Poincar\'e algebra). Due to this high degree of symmetry several 
expressions ---like the supersymmetry transformation laws for the fields---
have a relatively simple form. In a second step one reduces the symmetry 
algebra to the 
super Poincar\'e algebra by implementing a partial gauge choice. Moreover 
one eliminates several auxiliary fields. In doing so the supersymmetry 
transformations in general acquire a more complicated form, but in any case
they can be obtained by a number of well-defined algorithmical steps.

What makes the perturbative heterotic string so interesting from an $N=2$ 
supergravity point of view is that it gives us two different examples of 
theories where the direct application of the superconformal ideas is 
problematic. The first example arises when one goes to the stringy vector 
formulation for which a prepotential --- which is a crucial ingredient in 
any off-shell superconformal vector multiplet theory---  doesn't exist 
\cite{CDFVP}.\footnote{This case can be treated in an elegant way if one goes 
on-shell, where it is possible to write down an action \cite{geometr} in 
terms of symplectic sections rather then the prepotential itself. In the 
stringy basis these symplectic sections remain well-defined even though 
they are no longer based on an underlying prepotential.} The second example
concerns the heterotic antisymmetric tensor theory, which, as we already said,
escaped any direct superconformal treatment along the lines of \cite{wij}. 

Although we know in advance that somehow the off-shell superconformal 
framework must break down when we dualise towards these 
``problematic'' theories, we perform our computations in a superconformal 
setting. First of all this enables us to verify which ingredient of the
standard superconformal framework is incompatible with the dualisation 
procedure. We find that the duality transformations that are used to obtain
the two problematic theories both affect
the Weyl multiplet, such that the latter is no longer realised 
off-shell\footnote{Note that this departure from off-shellness does not occur 
in the cases decribed in \cite{wij} because 
there the Weyl multiplet is never touched.}. In this sense the
nonexistence of a conventional superconformal description for both the 
stringy vector multiplet version and the antisymmetric tensor version of 
the heterotic effective action has a common origin. There is no problem in 
preserving the
various superconformal symmetries during the duality transformations. This
is important from a technical point of view, because it implies that we can 
determine the supersymmetry transformation laws for the stringy vectors and
antisymmetric tensor field before having to impose the conventional 
superconformal gauge choices and even before eliminating the auxiliary 
fields. In this way the necessary computations can be kept relatively simple. 

One of the beautiful aspects of the superconformal setup is that it 
allows for some flexibility in the Poincar\'e reduction. We can benefit 
from this flexibility by choosing a dilatation gauge which immediately 
leads to a string frame Poincar\'e action. It is satisfactory to see that 
from a purely supergravity point of view the string frame is selected 
as a very natural one. It can be defined as the only dilatation gauge
which makes the supersymmetry transformations of the stringy vector fields
dilaton independent. Once this gauge has been chosen one may verify that 
the antisymmetric tensor $\cB_{\mu\nu}$ does not transform into the 
dilaton or dilatini, which means that in the string frame these fields 
don't combine into a vector-tensor supermultiplet. What happens instead is
that the dilaton and the dilatini effectively become part of the 
compensating hypermultiplet, whereas the antisymmetric tensor becomes part
of the gravitational multiplet. Of course one can go to the 
Einstein frame,
by applying an interpolating dilatation and $S$-supersymmetry 
transformation. The supersymmetry variation of $\cB_{\mu\nu}$ is not 
altered by this step because $\cB_{\mu\nu}$ is inert under the 
interpolating transformations. On the other hand the Einstein gravitinos 
are dilatino dependent functions of the original string gravitinos, so 
when $\d \cB_{\mu\nu}$ is written in terms of the Einstein frame variables
 a (spurious) dilatino dependence gets induced. One might interprete the 
resulting Einstein frame configuration as an on-shell heterotic 
vector-tensor multiplet, but this configuration is clearly not selected by
string theory.

\setcounter{equation}{0}
\section{N=2 superconformal supergravity coupled to vector and
hypermultiplets}

In this section we discuss some basic facts concerning $d=4$ $N=2$
superconformal supergravity and its couplings to vector and
hypermultiplets. Our discussion will be rather brief, as we intend to use
the standard superconformal vector and hypermultiplet theories merely as a
starting point for our construction of the antisymmetric tensor version of
the low-energy effective action for heterotic strings. Symplectic 
transformations are discussed in some more detail though, because as far as 
we know they never received a full treatment in the superconformal framework. 
The interested reader can find more details about the
superconformal approach to $N=2$ supergravity in the original articles
\cite{tensorcal,dWVP,dWLVP}. For more recent texts we refer to 
\cite{VPLonden} and also to the last article of \cite{wij} 
(which contains a comprehensive list of conventions).

The Weyl multiplet is a central object in the superconformal multiplet
calculus as it contains the gravitational degrees of freedom. It consists
of the vierbeins ${e_\mu}^a$, the gravitinos $\psi_\mu^i$, the gauge fields
$b_\mu, A_\mu, \Vv_\mu{}^i_{\, j}$ --- which gauge dilatations, chiral
$U(1)$ and $SU(2)$ transformations---and a set of matter fields: a selfdual
tensor $T_{\mu\nu\, ij}^+$ which is antisymmetric in its $SU(2)$ indices, a
real scalar $D$ and a doublet of chiral fermions $\chi^i$. As such the Weyl
multiplet forms the basic representation of (a deformed version of) the
superconformal algebra, which closes off-shell. This algebra consists of
general coordinate, local Lorentz, chiral $U(1)$ and $SU(2)$
transformations, dilatations, special conformal and $Q$- and
$S$-supersymmetry transformations. For future reference we list the
transformation laws for the vierbeins and the gravitinos
\brr \d e_\mu\,^a &=&
     \bpl \bar{\e}^i\g^a\psi_{\mu i}+h.c.\bpr - \Lambda_D e_\mu\,^a
     \,, \nonumber\\
     \d\psi_\mu^i &=& 2{\cal D}_\mu\e^i
     -\ft{1}{4} \s\cdot T^{-\, ij}\g_\mu\e_j
     -\g_\mu\eta^i - \bpl\ft{1}{2} \Lambda_D + \ft{i}{2} \Lambda_{\rm U(1)}
     \bpr \psi_\mu^i \, .
\err
Here $\e^i, \eta^i, \Lambda_D$ and $\Lambda_{\rm U(1)}$ are parameters for
$Q$- and $S$-supersymmetry transformations, dilatations and chiral $U(1)$
transformations respectively. General coordinate, local Lorentz and $SU(2)$
transformations (with parameter $\Lambda_i{}^j$) are not explicitly given
because these can automatically be inferred from the index structure 
of the fields that are being transformed. Here and in what follows a 
derivative $\Dd_\mu$ stands for a covariant derivative with respect to 
local Lorentz, dilatation, chiral $U(1)$, $SU(2)$ and gauge 
transformations. Since the parameter $\e^i$ carries non-trivial Weyl and 
chiral $U(1)$ weights one has that
\bq {\cal D}_\mu\e^i = \bpl \der_\mu - \ft{1}{2} \omega_\mu^{ab} \s_{ab}
    +\ft{1}{2} b_\mu +\ft{i}{2} A_\mu \bpr \e^i + \ft{1}{2} \Vv_\mu{}^i{}_j
\e^j \, . 
\eq
Here $\omega_\mu^{ab}$ is the (dependent) gauge field for local Lorentz 
transformations. 

Next one
introduces a set of vector multiplets (labeled by an index $I$) which can
be consistently coupled to the Weyl multiplet. Each vector multiplet
contains a complex scalar $X^I$, a vector gauge field $W_\mu^I$, a real
$SU(2)$ triplet of scalars $Y^I_{ij}$ and a doublet of chiral gauginos
$\O_i^I$. The lagrangian describing the most general couplings of vector
multiplets to superconformal supergravity was given in \cite{dWLVP} in
formula (3.9). We will use this lagrangian several times in what follows
and denote it by $e^{-1} \Ll_{\rm vector}$. In order to clarify our
notations we list the bosonic terms:
\brr e^{-1} \Ll_{\rm vector} &=& XN\bX \bpl \ft{1}{6} R + D  \bpr
     + N_{IJ} \Dd_\mu X^I \Dd^\mu \bX^J - \ft{1}{8} N_{IJ} Y^I_{ij}
     Y^{ij J} \nonumber\\
     &&- \ft{i}{8} \bF_{IJ} F_{\mu\nu}^{+ I} F^{+\mu\nu\, J}
     - \ft{1}{8} XN_I F_{\mu\nu}^{+ I} T^{+\mu\nu}_{ij} \ve^{ij}
     + \ft{1}{64} XNX \bpl T^{+}_{\mu\nu ij} \ve^{ij} \bpr^2 + h.c.
     \nonumber\\
     &&+ \mbox{fermionic terms} \, .
\label{lagr}
\err
This lagrangian depends on a holomorphic prepotential $F(X)$ which is
homogeneous of second degree in the $X^I$'s.\footnote{Note that compared to
\cite{dWLVP} we have rescaled the prepotential $F(X) \mapsto 2i F(X)$ as is
common practice in the recent literature. We also changed the sign of the
Ricci scalar $R$ such that under finite Weyl rescalings
\brr g_{\mu\nu}{}^{\prime} &=& e^{-2\Lambda_D} g_{\mu\nu} \nonumber\\
     e^{-2\Lambda_D} R^{\prime} &=& R + 6 \,\Box \Lambda_D - 6 \,\der_\mu
     \Lambda_D \der^\mu \Lambda_D  \, . \nonumber
\err
} Derivatives of the prepotential
with respect to the scalars $X^I$ are denoted by $F_I, F_{IJ}, \cdots$, and
$N_{IJ} \mathrel{\mathop{=}\limits^{\rm def}} - \mbox{Im}\, F_{IJ}$.
Sometimes we omit contracted $I$ indices, e.g. $\bX N_I = \bX^J N_{JI}$.
Furthermore we use abelian vectors only. This is a necessity for those
vectors we want to dualise at some point. Although we could keep (some of)
the others non-abelian we choose not to do so for simplicity. The field
strengths, their duals and (anti)selfdual parts are given by
\bq F_{\mu\nu}^I = 2 \der_{[ \mu} W_{\nu ]}^I \qquad \qquad
    \tilde{F}^{\mu\nu \,I} = \ft{1}{2} e^{-1} \ve^{\mu\nu\l\s} F_{\l\s}^I
    \qquad \qquad F_{\mu\nu}^{\pm I} = \ft{1}{2} \bpl F_{\mu\nu}^I\pm
    \tilde{F}_{\mu\nu}^I\bpr \, .
\eq
We take $\ve^{0123}=i$ such that complex conjugation interchanges
$F_{\mu\nu}^{+I}$ with $F_{\mu\nu}^{-I}$. The vector multiplet fields
transform as follows:
\brr \d X^I &=& \bar{\e}^i\O_i^{\,I} + \bpl \Lambda_D -i\Lambda_{\rm U(1)}
     \bpr X^I \nonumber\\
     \d\O_i^{\,I} &=& 2\Dslash X^{I}\e_i + \ve_{ij}\s^{\mu\nu} \e^j
     \bpl \hF_{\mu\nu}^{-I} - \ft{1}{4} \bX^I T_{\mu\nu}^{-\, kl}
     \ve_{kl} \bpr  +Y_{ij}^{\,I}\e^j +2X^{I}\eta_i +\bpl \ft{3}{2}
     \Lambda_D - \ft{i}{2}\Lambda_{\rm U(1)} \bpr \O_i^{\,I}
     \nonumber\\
     \d W_\mu^I &=& \bpl \bar{\e}_i\g_\mu\O_j^{\,I} \ve^{ij}
     +2\bX^I  \bar{\e}^i\psi_\mu^j \ve_{ij} +h.c. \bpr + \der_\mu \theta^I
     \nonumber\\
     \d Y_{ij}^{\,I} &=& 2\bar{\e}_{(i}\Dslash\O_{j)}^I
     +2\ve_{ik}\ve_{jl}\bar{\e}^{(k}\Dslash\O^{l)\,I} +
     2\Lambda_D Y_{ij}^{\,I}  \, .
\label{vrules}
\err
The derivatives $D_\mu$ are covariant with respect to all the
superconformal (and possibly also gauge) symmetries. The covariant field
strengths for the vectors are given by
\bq \hF_{\mu\nu}^I = 2 \der_{[\mu } W^I_{\nu ]}
    +\bpl  \bO^{iI} \g_{[ \mu }\psi_{\nu ]}^j \ve_{ij}
    - \bX^I \bpsi^i_\mu \psi_\nu^j \ve_{ij} +h.c. \bpr \, .
\label{covstrength}
\eq
Finally one adds a set of hypermultiplets consisting of $r$ quaternions
$A_i{}^{\a}$ and $2r$ chiral fermions $\zeta^{\a}$, where $\a = 1, \cdots,
2r$. The scalar fields $A_i{}^{\a}$ satisfy a reality constraint given in
\cite{dWLVP} which implies that they describe $4r$ real degrees of
freedom. In order to obtain a fully off-shell description for the
hypermultiplets, one must add a set of $4r$ auxiliary scalars. These
auxiliary degrees of freedom completely decouple from the other fields, so
we choose to integrate them out right from the start. Having done so we get
the following transformation rules for the hypermultiplets
\brr \d A_i{}^\a &=& 2\bar{\zeta}^\a\e_i
     +2\rho^{\a\b}\ve_{ij}\bar{\zeta}_\b\e^j + \Lambda_D  A_i{}^\a \nonumber\\
     \d\zeta^\a &=& \Dslash A_i{}^\a\e^i
     +A_i{}^\a\eta^i +\bpl \ft{3}{2}
     \Lambda_D - \ft{i}{2}\Lambda_{\rm U(1)} \bpr\zeta^\a \,.
\label{hrules}\err
The superconformal action describing the hypermultiplet couplings to the Weyl
multiplet is given in \cite{dWLVP} in formula (3.29). We list its
bosonic part:
\brr e^{-1} \Ll_{\rm hyper} &=&
     \ft{1}{2} A_i{}^\a A^i{}_\b d_\a{}^\b\bpl -\ft{1}{3} R + D \bpr
     - \Dd_\mu A_i{}^\a \Dd^\mu A^i{}_\b  d_\a{}^\b \nonumber\\
     &&+ \mbox{fermionic terms} \, .
\label{lagrhyper}
\err
For conventions concerning the constant matrices $\rho^{\a\b}$ and
$d_\a{}^\b$ we refer to \cite{dWLVP}.

\subsection{Symplectic transformations}

The superconformal vector multiplet theories we just described can be
acted upon with symplectic transformations. In general these transformations
relate a given theory (characterised by a prepotential $F(X)$) to other
vector multiplet theories which are classically equivalent to the first
one, in the sense that their equations of motion and Bianchi identities are
transformed into eachother.

Symplectic transformations can be introduced in the following way. First
of all one requires that they act linearly on the equations of motion and
the Bianchi identities that are satisfied by the field strengths for the
vectors \cite{FSZ}:
\bq \begin{array}{ll} \der^\mu \bpl F_{\mu\nu}^{- I}
    - F_{\mu\nu}^{+ I} \bpr = 0 \hspace{2cm} &\mbox{Bianchi identities}
    \\[2mm]
    \der^\mu \bpl G_{\mu\nu\, I}^-  - G_{\mu\nu\, I}^+ \bpr = 0
    &\mbox{Equations of motion} \\[2mm]
    G_{\mu\nu\, I}^- \mathrel{\mathop{=}\limits^{\rm def}}
    -4i\frac{\textstyle \d \, e^{-1} \Ll_{\rm vector}}{\textstyle
    \d F^{\mu\nu - I}}
    \end{array}
\label{eomandBian}
\eq
This can be accomplished by putting $F_{\mu\nu}^{- I}$ and $G_{\mu\nu\,
I}^-$ into a vector transforming as
\bq \left(\begin{array}{c} F_{\mu\nu}^{- I} \\[1mm] G_{\mu\nu\, I}^-
    \end{array}\right) \longrightarrow
    \left(\begin{array}{c} F_{\mu\nu}^{- I}{}' \\[1mm] G_{\mu\nu\, I}^-{}'
    \end{array}\right) =
    \left(\begin{array}{cc} U^I{}_J &Z^{IJ} \\[1mm]  W_{IJ} & V_I{}^J
    \end{array}\right)
    \left(\begin{array}{c} F_{\mu\nu}^{- J} \\[1mm] G_{\mu\nu\, J}^-
    \end{array}\right) \, ,
\label{symeom}
\eq
where $U^I{}_J, Z^{IJ}, W_{IJ}, V_I{}^J$ are constant real matrices. These
matrices cannot be chosen arbitrarily because one must ensure that the
$G_{\mu\nu\, I}^-{}'$ result from varying the new vector multiplet theory
with respect to the new field strengths $F_{\mu\nu}^{- I}{}'$. This implies
\cite{dWVP} that the transformation must be
symplectic, i.e.
\bq \left(\begin{array}{cc}  U^T &W^T \\[1mm]  Z^T & V^T \end{array}\right)
    \left(\begin{array}{cc}  0 &\unity \\[1mm]  -\unity & 0
    \end{array}\right)
    \left(\begin{array}{cc}  U &Z \\[1mm]  W & V \end{array}\right) =
    \left(\begin{array}{cc}  0 &\unity \\[1mm]  -\unity & 0
    \end{array}\right) \label{defsympl}
\eq
and that the scalars $X^I$ and the $F_I$ must form a symplectic vector too:
\bq \left(\!\begin{array}{c} X^I \\[1mm] F_I
    \end{array}\!\right) \longrightarrow
    \left(\!\!\begin{array}{c} X^I{}' \\[1mm] F_I{}'
    \end{array}\!\!\right) =
    \left(\begin{array}{cc} U^I{}_J &Z^{IJ} \\[1mm] W_{IJ} & V_I{}^J
    \end{array}\right)
    \left(\!\begin{array}{c} X^J \\[1mm] F_J
    \end{array}\!\right) \, .
\label{basicvec}
\eq
In the present superconformal setup one has to
impose an additional restriction on the concept of symplectic
transformations. The transformed theory only makes sense as a
superconformal theory when it is based on a new prepotential, say $F'(X')$.
But this presupposes that all the transformed scalars $X^I{}'$ are
independent variables, which excludes some a priori interesting symplectic
transformations\footnote{It is well-known that one can circumvent this last
restriction if one goes on-shell. The point is that the Poincar\'e theories
for $N=2$ vector multiplets can be formulated in terms of symplectic
sections $(X^I, F_I)$ \cite{geometr}, which need not be derived
from a prepotential. The restriction on the admissable symplectic
transformations then evaporates accordingly \cite{CDFVP}.}. The symplectic
transformation which transforms the {\it perturbative} heterotic $STU$
theory into its stringy analog is a particular example of a transformation
which leads to new scalars which are not all independent. In section 3.2 
we verify in which sense this forces us to
exceed the off-shell superconformal framework\footnote{In fact we will show
that one can stay quite close to the original superconformal setup. It
suffices to go partly on-shell by eliminating the $T^+_{\mu\nu\, ij}$
auxiliary field.}.

Note that only a subset of the full symplectic group maps the original
theory onto itself. These so-called duality invariances \cite{dWVP} are
characterised by the fact that $F'(X') = F(X')$\footnote{Remark that $F(X)$
is {\it not} an invariant function because $F'(X')\neq F(X)$.}. This implies
that the $F_I{}' = W_{IJ} X^J + V_I{}^J F_J(X)$ are simply given
by $F_I(X')$. Sometimes the duality invariances are also called proper
symmetries, while the other symplectic transformations which really change
the form of the prepotential are called symplectic reparametrisations
\cite{CFG} or pseudo-symmetries \cite{pseudo}.

Let us now proceed with an explicit description of the symplectic
properties of the lagrangian (\ref{lagr}) including the fermionic terms
and the Weyl multiplet auxiliaries. Of course many of these symplectic
properties overlap with those that have been studied previously in the
Poincar\'e case \cite{dWVP,dWVVP,CDFVP}, so we will concentrate on the
ingredients that are specific to the superconformal setup. See also
\cite{chiral} for a (purely bosonic)
treatment of symplectic transformations in the presence of a Weyl multiplet
background.

The symplectic vectors that are relevant for \eqn{lagr} can be derived
starting from the basic symplectic vector $(X^I,  F_I)$ by requiring
consistency with supersymmetry. If one computes succesive supersymmetry
variations of the basic symplectic vector, one gets the following
results\footnote{For a related discussion, see \cite{CRTVP} in
which symplectic transformations are defined on entire $N=2$ chiral
superfields.}:
\brr &&\begin{array}{lcl}
     \d_Q \left(\!\begin{array}{c} X^I \\[1mm]
     F_I \end{array}\!\right) \!\!&=&\!\!   \bar{\e}^i
     \left( \!\!\! \begin{array}{c} \O_i^{\,I} \\[1mm]
     F_{IJ} \O_i^{\,J}
     \end{array} \!\!\! \right) \nonumber\\
     \d_Q \left(\!\!\! \begin{array}{c} \O_i^{\,I} \\[1mm] F_{IJ}
     \O_i^{\,J} \end{array}\!\!\!\right)\!\! &=& \!\!2\g^\mu\e_i \,D_\mu
     \left(\!\begin{array}{c} X^I \\[1mm] F_I \end{array}\!\right)
     + \ve_{ij} \s^{\mu\nu} \e^j
     \left\{ \left(\begin{array}{c} \hF_{\mu\nu}^{-I} \\[1mm]
     \hG_{\mu\nu I}^- \end{array}\right)
     - \frac{1}{4} T_{\mu\nu}^{-\, kl} \ve_{kl}
     \left(\begin{array}{c} \bX^I \\[1mm] \bF_I \end{array}\right)
     \right\} \nonumber\\
     && + \e^j \left(\!\begin{array}{c} Y_{ij}^{\,I} \\[1mm]
     Z_{ij\, I}\end{array}\!\right)
     \end{array} \nonumber\\[3mm]
     && \hG_{\mu\nu I}^- \mathrel{\mathop{=}\limits^{\rm def}}
     F_{IJ} \hF_{\mu\nu}^{-J}
     + \ft{i}{2} \bX N_I T_{\mu\nu}^{- \, ij} \ve_{ij}
     - \ft{1}{4} F_{IJK}  \bO_i^J \s_{\mu\nu}\O_j^K \ve^{ij} \nonumber\\
     && Z_{ij\, I}\,\, \mathrel{\mathop{=}\limits^{\rm def}}  F_{IJ}
     Y_{ij}^J - \ft{1}{2} F_{IJK} \bO_{(i}^J \O_{j)}^K \,\,\, .
\label{symplvec}
\err
The expressions $(\O_i^{\,I}, F_{IJ} \O_i^{\,J})$ and $(\hF_{\mu\nu}^{-I} ,
\hG_{\mu\nu\, I}^-)$ automatically define new consistent symplectic
vectors. Note that $(\hF_{\mu\nu}^{-I}, \hG_{\mu\nu\, I}^-)$ is equivalent
to the vector $(F_{\mu\nu}^{-I}, G_{\mu\nu\, I}^-)$ we introduced in 
\eqn{eomandBian} and \eqn{symeom}, because the difference between the two
is itself symplectic. However, the would-be vector $(Y_{ij}{}^I, Z_{ij\,
I})$ is not automatically consistent. The reason is that the auxiliary
fields $Y_{ij}{}^I$ satisfy a reality condition, viz. $Y_{ij}{}^I \ve^{jk}
= \ve_{ij} Y^{jk\,I}$, whereas the $Z_{ij\, I}$ apparently don't. But one
can verify that the $Z_{ij\, I}$ do satisfy a similar reality condition if
one imposes the $Y_{ij}{}^I$ equations of motion:
\bq N_{IJ} Y_{ij}{}^J - \ft{i}{4} F_{IJK} \bO_i^J \O_j^K
    + \ft{i}{4} \bF_{IJK} \bO^{kJ} \O^{lK} \ve_{ik} \ve_{jl} = 0 \, .
\label{Yeom}
\eq
We henceforth assume that these equations of motion have been
enforced. The $Y_{ij}{}^I$'s must then be viewed as particular dependent
expressions quadratic in the gauginos $\O_i^I$ or their complex conjugates.
Nevertheless we will find it useful to keep on using the shorthand
$Y_{ij}{}^I$ in what follows.

It is clear from equation \eqn{symplvec} that the symplectic vectors
$(X^I, F_I)$ and $(\O_i^{\,I},
F_{IJ} \O_i^{\,J})$ transform into other symplectic vectors under
supersymmetry. This fact holds in general although one has to be careful.
One may verify that the symplectic vector $(\hF_{\mu\nu}^{-I},
\hG_{\mu\nu\, I}^-)$ only transforms into other symplectic vectors provided
one uses the equations of motion for the gauginos $\O_i^I$. At this point,
however, we prefer not to impose the latter equations of motion, because
they are not needed in order to guarantee the consistency of the symplectic
vectors we just defined. Therefore we simply use the symplectic vectors of
\eqn{symplvec} as they stand. The resulting symplectic transformation rules 
for the vector multiplet fields make sense at a lagrangian level, and the 
lagrangian \eqn{lagr} --- with dependent $Y_{ij}{}^I$, but all
the other Weyl multiplet auxiliaries still present as independent degrees
of freedom--- turns out to be symplectically invariant, up to a familiar
${\rm Im}( F_{\mu\nu}^{+I} G^{\mu\nu\, +}_I)$ term. 

When checking the behaviour of the lagrangian \eqn{lagr} under symplectic
transformations it suffices to concentrate on those terms that would vanish
if the auxiliary fields were eliminated and the superconformal gauge
choices were imposed. The symplectic properties of the other terms are
known in advance as they are not changed by the transition to the
Poincar\'e theory for which a complete treatment has already appeared
elsewhere \cite{CDFVP,dWVP}. One finds that the vector multiplet lagrangian
can be written as
\bq e^{-1} \Ll_{\rm vector} = -\ft{i}{8}   F_{\mu\nu}^{+I}
    G^{\mu\nu\, +}_I +h.c. + \mbox{invariant terms} \, ,
\label{symact}
\eq
which is completely analogous to the Poincar\'e result. In principle there
could have been additional non-invariant terms, which then would have to
vanish in the Poincar\'e limit, but such terms do not appear. As an 
illustration, we treat the $T_{\mu\nu ij}^+$ terms explicitly and show 
that they can
be nicely absorbed into symplectic invariant combinations. We start with
the invariant
\brr &&\ft{i}{32}  \bl X^I \hG_{\mu\nu\, I}^+ - F_I \,
     \hF^{+ I}_{\mu\nu} \br T^{\mu\nu +}_{ij} \ve^{ij}+ h.c. \nonumber\\
     &&= \ft{1}{64}
     \bl XNX\, T^{+}_{\mu\nu kl} \ve^{kl} - 4 XN_I\, \hF_{\mu\nu}^{+ I}
     - \ft{i}{2} X^I \bF_{IJK}  \bO^{kJ} \s_{\mu\nu}\O^{lK} \ve_{kl} \br
     T^{\mu\nu +}_{ij} \ve^{ij}+ h.c.
\label{Tinv}
\err
which contains all the $\Oo (T_{\mu\nu ij}^+)^2$ terms of \eqn{lagr}. After
taking into account the terms linear in $T_{\mu\nu ij}^+$ which appear in
\eqn{Tinv} and in the $F_{\mu\nu}^{+I} G^{\mu\nu\, +}_I$ term of
\eqn{symact} one is left over with
\brr &&\ft{1}{24}  \bl XN_I \,\bO^{iI}\g_\mu \psi^j_\nu +  XN\bX\,
     \bpsi^i_\mu \psi^j_\nu \br T^{\mu\nu +}_{ij} + h.c. \nonumber\\
     && -\ft{1}{16}  XN_I \bl  \bO_i^I\g_\mu \psi_{\nu j} \ve^{ij} -X^I
     \bpsi_{\mu i} \psi_{\nu j} \ve^{ij} \br T^{\mu\nu +}_{kl} \ve^{kl}
     + h.c. \nonumber\\
     && - \ft{i}{64}  X^I \bF_{IJK}  \bO^{iJ}
     \s_{\mu\nu}\O^{jK} T^{\mu\nu +}_{ij} + h.c.
\label{remain}
\err
The first line of \eqn{remain} is invariant by itself. The second line
conspires with suitable $T^+_{\mu\nu ij}$ independent terms in \eqn{lagr}
to form the invariant
\bq -\ft{i}{8} \bl G_I^{\mu\nu+} \bpl \bO_i^I\g_\mu \psi_{\nu j}
    \ve^{ij} -X^I \bpsi_{\mu i} \psi_{\nu j} \ve^{ij} \bpr
    - F^{\mu\nu + I} \bpl F_{IJ} \bO_i^J\g_\mu \psi_{\nu j}
    \ve^{ij} -  F_I\, \bpsi_{\mu i} \psi_{\nu j} \ve^{ij} \bpr
    \br +h.c.
\label{invar2}
\eq
One can check that \eqn{Tinv}, \eqn{invar2} and the
$F_{\mu\nu}^{+I} G^{\mu\nu\, +}_I$ term of \eqn{symact} already contain all
the $F_{\mu\nu}^I$ terms of the lagrangian apart from
\bq \ft{i}{32} F^{\mu\nu +I} \bF_{IJK}  \bO^{iJ}
    \s_{\mu\nu}\O^{jK} \ve_{ij} + h.c.
\label{last}
\eq
But the sum of \eqn{last} and the last line of \eqn{remain} transforms into
itself up to 4-fermi terms (which can be further analysed in the same way as
in the Poincar\'e case).

To finish our discussion of symplectic transformations we introduce a
useful piece of terminology \cite{CDFVP}. A given symplectic transformation
is said to be of the ``semi-classical" type when $Z^{IJ}=0$ (in a well
chosen symplectic basis). In that case the full set of Bianchi
identities is left invariant, which implies that the vectors transform
linearly into themselves. For $Z^{IJ} = 0$ one makes a further distinction
between the $W_{IJ} = 0$ case which is called ``classical", and the
genuine ``semi-classical" case $W_{IJ} \neq 0$. It follows from
\eqn{symeom},
\eqn{defsympl} and \eqn{symact} that the lagrangian \eqn{lagr} is left
invariant under semi-classical transformations up to a topological term:
\bq e^{-1} \Ll_{\rm vector}
    \mathrel{\mathop{\longrightarrow}\limits^{\rm semi-class.}}
    e^{-1} \Ll_{\rm vector}
    -\ft{i}{8}  F_{\mu\nu}^I [ U^T W]_{IJ} {\tilde F}^{\mu\nu J}
    \, .
\eq

\subsection{The transition to the Poincar\'e theory}

The off-shell theories for $N=2$ vector and hypermultiplets coupled to
superconformal supergravity contain some redundant variables which do not
describe true physical degrees of freedom. These redundant variables can be
eliminated by going on-shell and by reducing the superconformal symmetry
algebra to the super Poincar\'e algebra. As has been explained in
\cite{dWLVP} there exists a well-defined procedure to perform this step,
and we now recall its most important ingredients.

As a starting point one considers the sum of the vector multiplet action
\eqn{lagr} and the hypermultiplet action \eqn{lagrhyper}. The total number
of hypermultiplets one introduces is equal to $r = (N_h + 1)$, where $N_h$
denotes the number of physical hypermultiplets one wants to obtain in the
final theory. The extra hypermultiplet is a so-called compensating
multiplet which plays an important r\^ole in the whole Poincar\'e reduction.
In a moment we will sketch how the degrees of freedom of this compensating
hypermultiplet can be eliminated. In the vector multiplet sector there are
some compensating degrees of freedom too: one complex scalar and its
fermionic partner are unphysical and as such they will be removed in the
process of going to the Poincar\'e theory. This implies that the number $I$
always runs over $(N_v + 1)$ values, where $N_v$ counts the (complex)
dimension of the vector multiplet moduli space.

The Weyl multiplet fields $D , \chi_i , A_\mu , \Vv_\mu{}^i_{\, j}$, and
$T_{\mu\nu\, ij}^+$ are auxiliary fields. $D$ and $\chi_i$ are
Lagrange multipliers which enforce the following constraints on the
matter sector of the theory:
\brr XN\bX\, + \ft{1}{2} A_i{}^\a A^i{}_\b d_\a{}^\b &=& 0 \nonumber\\
     \bX N_I \O_i^I + 2 A_i{}^\a \zeta_\b d_\a{}^\b &=& 0 \, .
\label{cons}
\err
One can use these constraints to eliminate several component fields of the
compensating hypermultiplet, i.e. one real component of the scalar and all
the components of the associated fermion. The auxiliary fields
$A_\mu , \Vv_\mu{}^i_{\, j}$, and $T_{\mu\nu\, ij}^+$ appear quadratically
in the action so they can be solved for by imposing their own equations of
motion. In the sequel we will explicitly need the $A_\mu , \Vv_\mu{}^i_{\,
j}$ and $T_{\mu\nu\, ij}^+$ equations of motion which are given by
\brr XN\bX\, A_\mu &=& \bpl \ft{i}{2} \bX N_I \hat{\der}_\mu X^I+h.c.
     \bpr - \ft{i}{8}  N_{IJ} \bO^{iI}\g_\mu \O_i^J - i \bar{\zeta}^\a
     \g_\mu \zeta_\b d_\a{}^\b \nonumber\\
     -\ft{1}{2} A_k{}^\a A^k{}_\b d_\a{}^\b \, \Vv_\mu{}_i{}^j
     &=&  -\bpl A_i{}^\a \hat{\der}_\mu A^j{}_\b d_\a{}^\b - A^j{}_\b
     \hat{\der}_\mu A_i{}^\a d_\a{}^\b \bpr \nonumber\\
     &&- \ft{1}{2}  N_{IJ} \bO^I_i\g_\mu \O^{jJ}
     + \ft{1}{4} \d_i{}^j N_{IJ} \bO^I_k\g_\mu \O^{kJ}  \nonumber\\
     XNX\, T_{\mu\nu \, ij}^+ \ve^{ij} &=&  4XN_I \hF_{\mu\nu}^{+I}
     +\ft{i}{2} X^I \bF_{IJK}  \bO^{iJ}\s_{\mu\nu} \O^{jK}\ve_{ij}
     - 8\, d_\a{}^\b \rho_{\b\g}\,\bar{\zeta}^\a \s_{\mu\nu} \zeta^\g 
     \, . 
\label{Teomgen}
\err
In order to derive the equation \eqn{Teomgen} we used the constraints
\eqn{cons}. We also neglected any $b_\mu$ dependence, because we will put 
$b_\mu = 0$ in a moment. By $\hat{\der}_\mu$ we mean a derivative which is
covariant with respect to local Lorentz, $Q$-supersymmetry and gauge
transformations\footnote{Our convention for $\hat{\der}_\mu$ differs from
the one given in \cite{dWLVP} in that our $\hat{\der}_\mu$ is covariant
with respect to $Q$-supersymmetry as well.}. One easily verifies that the
equations \eqn{Teomgen} are symplectically invariant, see for instance 
\eqn{Tinv}.

The superconformal algebra can be broken down to the super Poincar\'e
algebra by imposing several gauge choices. First one breaks the special
conformal symmetry by putting the dilatation gauge field $b_\mu = 0$.
Next one uses the dilatation symmetry to bring the $R$ term in the action
into a conventional form. It is common practice in the $N=2$ supergravity
literature to go to the Einstein frame for the metric, i.e. $e^{-1} \Ll =
\frac{1}{2} R$ + more. Later on in the heterotic string case (section 5.4)
we will not follow this common practice, but rather choose an alternative
gauge which immediately leads to the string frame. Nevertheless we present
the standard Einstein frame results here, so that the reader may compare
to them
in section 5.4. Given the fact that the $D$ equation of motion \eqn{cons}
has already been imposed, one finds that the dilatation gauge choice
leading to the Einstein frame reads
\bq XN\bX = 1    \, .
\label{Dchoice}
\eq
The corresponding standard $S$-supersymmetry gauge choice reads
\bq XN_I\O^{iI} = 0  \, .
\label{Schoice}
\eq
This $S$-gauge choice is chosen such that the dilatation gauge choice
\eqn{Dchoice} is invariant under $Q$-supersymmetry transformations.
Moreover one may verify that \eqn{Schoice} removes many of the mixed
gravitino-gaugino propagators. It is important to realise that the
condition \eqn{Schoice} itself is not $Q$-supersymmetric invariant. This
shows that the Poincar\'e sypersymmetry transformations ---which by
definition leave the various gauge choices invariant--- are in fact
composed out of a $Q$-supersymmetry part and a compensating
$S$-supersymmetry part:
\brr \d (\e) &=& \d_Q (\e) + \d_S (\eta(\e)) \nonumber\\
     \eta_i (\e) &=&  \g^\mu \e_j \bl \ft{1}{4} N_{IJ} \bpl
     \bO{}^{jI} \g_\mu \O^J_i -\ft{1}{2} \d^j{}_i \bO{}^{kI} \g_\mu \O^J_k
     \bpr \nonumber\\
     &&- \d^j{}_i \bar{\zeta}^\a \g_\mu \zeta_\b d_\a{}^\b \br
     - \ve_{ij} \s^{\mu\nu} \e^j d_\a{}^\b \rho^{\a\g} \bar{\zeta}_\b
     \s_{\mu\nu} \zeta_\g\, .
\err
In order to finish the whole Poincar\'e reduction one has to fix the chiral
$U(1)$ and $SU(2)$ symmetries. The $U(1)$ gauge freedom can be used to
further restrict the scalar fields $X^I$. If one introduces special
coordinates
\bq
Z^I \mathrel{\mathop{=}\limits^{\rm def}} -i \frac{X^I}{X^0}
\eq
one derives from eq.\eqn{Dchoice} that
\bq
|X^0|^2 = (ZN\bZ)^{-1} \, ,
\eq
so in this case the dilatation gauge choice has effectively fixed the
length of the scalar $X^0$. One then uses the $U(1)$ symmetry to fix the
phase of $X^0$ as well. A convenient choice is
\bq X^I = (ZN\bZ)^{\textstyle -\frac{1}{2}} Z^I \, .
\label{U1fix}
\eq
This expresses the (dependent) scalars $X^I$ in terms of the (independent)
$Z^I$. It should be noted that the previous formula and the
subsequent ones are also valid when the $Z^I$ are not just special
coordinates, but rather holomorphic sections $Z^I(z^\Aa)$ of a symplectic
bundle over the special K\"ahler manifold ${\cal SK}$ which is defined by
the vector multiplet theory \cite{holsections}. The $z^\Aa$ with $\Aa=(1,
\cdots ,N_v)$ are arbitrary coordinates on ${\cal SK}$.
The relation \eqn{U1fix} is not $Q$-supersymmetry invariant, which implies
that a
compensating $U(1)$ transformation must be included in order to obtain the
correct Poincar\'e supersymmetry transformation rules. The precise form of
this compensating transformation can be deduced from the fact that
\brr \d (\e) X^I &=& \bar{\e}^i\O_i^{\,I} -i\Lambda_{\rm U(1)} X^I
     \nonumber\\
     &=& (ZN\bZ)^{\textstyle -\frac{1}{2}} \bar{\e}^i\Lambda_i^{\,I}
     - \ft{1}{2} X^I \bpl \frac{\bZ N_J}{ZN\bZ}\e^i \Lambda_i^J + h.c.
     \bpr \, ,
\label{condit}
\err
where the fermions $\Lambda_i^{\,I}$ are defined by
\bq
\d (\e) Z^I = \bar{\e}^i\Lambda_i^{\,I} \, .
\eq
From \eqn{condit} one immediately reads off that
\brr \Lambda_{U(1)} (\e) &=& \frac{i}{2} \frac{\bZ N_I}{ZN\bZ} \e^i
     \Lambda_i^I +h.c. \nonumber\\
     \O^I_i &=& (ZN\bZ)^{\textstyle -\frac{1}{2}} \left( \Lambda_i^I - Z^I
     \frac{\bZ N_J\Lambda_i^J}{ZN\bZ} \right) \, .
\err
Remark that the $Z^I$ (and thus any K\" ahler coordinates $z^\Aa$) are 
defined to be chiral fields, in the sense that they only transform into
fermions of a definite handedness under Poincar\'e
supersymmetry\footnote{As usual we use Weyl fermions for which the
position of the $SU(2)$ index also indicates the chirality.}. The scalars
$X^I$ are not chiral under Poincar\'e supersymmetry because of the
compensating $U(1)$ transformation. The $SU(2)$ symmetry, finally, can be
used to remove the last 3 degrees of freedom of the compensating 
hypermultiplet scalar. A similar reasoning as the one followed for the 
chiral $U(1)$ gauge fixing leads to\footnote{To
clarify our notation $A_i{}^j - h.c.;\mbox{traceless} = A_i{}^j - A^j{}_i
- \frac{1}{2}\d_i^j ( A_k{}^k - A^k{}_k)$ and $A^j{}_i = (A_j{}^i)^*$. }
\brr A_i{}^\a &=& \sqrt{\frac{-2}{C(B)}} \,\d_i{}^s B_s^\a
     \hspace{1cm} s = 1,2 \hspace{1cm} C(B) = B_s{}^\a B^s{}_\b d_\a{}^\b
     \nonumber\\
     \zeta^\a &=& \sqrt{\frac{-2}{C(B)}} \left( \xi^\a - B_s{}^\a
     \frac{2 B^s{}_\b\, d^\b{}_\g \xi^\g}{C(B)} \right) \nonumber\\
     \Lambda_i{}^j &=& 4 \bar{\e}_i \d^j_s B^s{}_\a\, d^\a{}_\b \xi^\b
     C(B)^{-1} - h.c.;\mbox{traceless}
\err
where the $\xi^\a$ are defined by
\bq \d B_s{}^\a = \bpl 2\bar{\xi}^\a\e_i
    +2\rho^{\a\b}\ve_{ij}\bar{\xi}_\b\e^j\bpr \d^i{}_s \, .
\eq

\setcounter{equation}{0}
\section{The vector multiplet effective action for perturbative heterotic
strings on $K_3\times T^2$}

In this section we concentrate on a particular subset of all possible
$N=2$ vector multiplet theories, namely those that arise from type II
strings on $K_3$-fibered Calabi-Yau manifolds, or from heterotic
strings on $K_3\times T^2$. We begin with type II strings because the
vector multiplet sector of their low-energy effective action is relatively
easy to analyse. This is so because the type II dilaton is contained in the
hypermultiplet sector of the theory\footnote{One must be careful with this
statement, though. As has been explained in \cite{BS} the type II dilaton
is naturally described by the real part of the sum of a compensating
$N=2$ vector and tensor multiplet. Only after imposing an Einstein frame
dilatation gauge choice the type II dilaton appears to be sitting in a
physical tensor multiplet which can be converted into a hypermultiplet for
practical reasons. So it is only in an Einstein frame that the type II
dilaton can be identified with one of the hypermoduli.}. The
fact that the vector multiplet sector and in particular the prepotential of
the type II effective action do not depend on the hypermoduli implies
that they are completely fixed at the tree level and don't receive any
perturbative or non-perturbative corrections. The prepotential which is
relevant for type IIA strings compactified on a generic Calabi-Yau manifold
$Y$ (and for type IIB strings on the mirror manifold $\tilde Y$) is given
by \cite{Ftype2}
\brr && F_{\rm type II} (X^0,X^\Aa)  = i (X^0)^2 f_{\rm type II} (Z^\Aa)
     \nonumber\\
     && f_{\rm type II} (Z^\Aa) = \frac{1}{6} d_{\Aa\Bb\Cc} Z^\Aa Z^\Bb
     Z^\Cc - \frac{\zeta (3)}{16\pi^3} \chi (Y) + \frac{1}{8\pi^3}
     \sum_{d_1, ..., d_{n+1}} n^r_{d_1, ..., d_{n+1}}\, Li_3
     \Big[ e^{\textstyle -2\pi d_\Aa Z^\Aa} \Big] \nonumber\\[-2mm]
     && Z^\Aa \mathrel{\mathop{=}\limits^{\rm def}} -i \frac{X^\Aa}
     {X^0}\, .
\label{prepII}
\err
Here the index $\Aa$ runs from $1, ..., N_v = n+1\, (n\geq 1)$. The
special coordinates $Z^\Aa$ are identified with the complexified K\"ahler
moduli of $Y$, the $d_{\Aa\Bb\Cc}$ are the Calabi-Yau triple intersection
numbers, $\chi(Y)$ is the Euler characteristic and the rational instanton
numbers $n^r_{d_1, ..., d_{n+1}}$ count the number of rational curves of
multi degree $(d_1,...,d_{n+1})$ on $Y$. When $Y$ is a $K_3$-fibration over
$\IP^1$ \cite{KLM,AL} there is one distinguished modulus, the K\"ahler
modulus of the $\IP^1$ base, which we call $S$. We may choose a basis such
that
\bq Z^\Aa = (S, Z^A) \hspace{2cm} A = 2, ..., n+1 \, .
\label{Ssplit}
\eq
The intersection numbers of the $K_3$-fibered Calabi-Yau manifold then
satisfy
\bq d_{111} = 0 \hspace{1cm} d_{11A} = 0 \hspace{1cm} d_{1AB} = \eta_{AB}
    \, ,
\label{dconstr}
\eq
where the $n\times n$ ``metric" $\eta_{AB}$ is of signature
$(1,n-1)$.\footnote{It should be mentioned that one may want to perform some
additional linear redefinitions of the K\"ahler moduli \eqn{Ssplit} and the
corresponding scalars $X^\a$. Consider e.g. the type II model based on the
Calabi-Yau space $WP_{1,1,2,8,12}(24)$ \cite{KV}. This model contains 3
K\"ahler moduli $Z^1 = S, Z^2, Z^3$ with intersection numbers $d_{123}=1,
d_{133}=d_{223}= 2, d_{233}= 4, d_{333}= 8$. It is advantageous to define
$Z^2 = T-U, Z^3 = U$ where $T$ and $U$ can be identified with the standard
moduli of the $T^2$ in the dual heterotic picture. In this way $\eta_{TU} =
1, \eta_{TT} = \eta_{UU} = 0$.} Notice that the index $I$ -- which runs
over all the vector multiplets --- takes the values $0, 1, A$. As it will
be relevant further on, we introduce a $(n+2) \times (n+2)$ metric
$\eta_{IJ}$ of signature $(2,n)$, which is defined as
\bq \eta_{IJ} = \left(\begin{array}{ccc} 0 & 1 & 0 \\ 1 & 0 & 0 \\
    0 & 0 & \eta_{AB} \end{array}\right) \, .
\label{SOmetric}
\eq

\subsection{Identification of the dilaton-axion complex}
Thanks to the type II-heterotic duality hypothesis -- which has been tested
very succesfully in many circumstances, see for instance
\cite{KV,FHSV,KLM,AL,dualtest}, the nice review \cite{LF} and
references therein --- one can view the prepotential implied by the
relations \eqn{prepII}, \eqn{Ssplit}, \eqn{dconstr} as a prepotential
describing heterotic strings on $K_3\times T^2$. Of course it is crucial to
make contact with the results which have been obtained in the past by
direct computations in the heterotic string picture \cite{dWKLL,AFGNT,HM}.
In order to do so, one identifies the modulus $S$ with the heterotic
dilaton-axion complex:
\bq S = -i \frac{X^1}{X^0}
    \mathrel{\mathop{=}\limits^{\rm def}} \phi -i a \, .
\eq
Here $\phi$ is a dilaton-like field which is closely related to the
heterotic string loop counting parameter, while $a$ is an axion-like
field\footnote{Henceforth we simply call $\phi$ the dilaton and $a$ the
axion, even though they differ from the true $SO(2,n)$ invariant dilaton
$\phi_{\rm inv}$ and axion $a_{\rm inv}$ which will be discussed later.}.
The heterotic prepotential thus satisfies the following expansion in the
string coupling constant $S$:
\bq  F_{\rm het} (X) = -\frac{1}{2} \frac{X^{1}}{X^0} \eta_{AB} X^A X^B
     + F^{(0)}(X^0,X^A) + \sum_{k=1}^{\infty}
     F^{(k)}(X^0,X^A)\,e^{\textstyle -2\pi k S} \, .
\label{expand}
\eq
At the tree level one recovers the well-known prepotential $-\frac{1}{2}
\frac{X^{1}}{X^0} \eta_{AB} X^A X^B$ corresponding to the special K\"ahler
manifold
\bq {SU(1,1)\over U(1)}\otimes {SO(2,n)\over SO(2)\times SO(n)}
\label{SKM}
\eq
which describes the classical vector moduli space for heterotic strings.
This manifold is the only special K\"ahler manifold having a direct product
structure \cite{FVP}. This reflects the fact that (in the Einstein frame)
the heterotic dilaton has no tree-level couplings to the other vector
moduli. When $n\geq 2$ the
tree-level prepotential can be brought into a standard form by performing
some linear redefinitions of the special coordinates $Z^A$ to find
\bq F_{\rm tree} = -\frac{1}{2}\frac{X^{1}}{X^0} \eta_{AB} X^A X^B
    = i (X^0)^2 \bl STU - S\sum_{i=4}^{n+1} \phi^i \phi^i \br \, .
\label{Ftree}
\eq
The fields $T$ and $U$ are the moduli of the $T^2$ and the $\phi^i$ are
possible Wilson line moduli. The $n=1$ case can be viewed as degenerate
case for which the difference $T-U$ has been frozen to a zero value. Due to
the result \eqn{Ftree} we say that the prepotential \eqn{expand} leads to
the heterotic vector multiplet effective action in the $STU$ basis.

The other terms in \eqn{expand} can be ``explained" by noting that they
are the only possible ones allowed by the quantised Peccei-Quinn symmetry
under which the heterotic theory is invariant. This symmetry maps
\bq S \mathrel{\mathop{\longrightarrow}\limits^{\rm P.Q.}} S -i c
    \qquad\qquad c\in\Zbar \, ,
\label{PQS}
\eq
and leaves the other moduli untouched. In order to fully understand the
consequences of this particular symmetry one needs some extra knowledge
which will be provided in section 4.1 where we study generic Peccei-Quinn
invariant models. At present it suffices to mention that in the case at
hand the prepotential has to transform as follows\footnote{This follows
from the equations \eqn{PQ} and \eqn{Wexplit} (appropiately applied to the
heterotic string example) and the fact that $F(X) = \frac{1}{2} X^I F_I$.
At first sight the reader might be surprised by the fact that the
prepotential is not invariant under the Peccei-Quinn symmetry even though
this symmetry corresponds to a duality invariance. The point is that
$F'(X') = F(X')$ but $\neq F(X)$. }
\bq F_{\rm het}(X)
\mathrel{\mathop{\longrightarrow}\limits^{\rm P.Q.}} F_{\rm het}(X)
- \ft{1}{2} c\,  \eta_{AB} X^A X^B \, .
\eq
Since the tree level part of the prepotential already saturates the latter
equation all the other terms must be Peccei-Quinn invariant by themselves
which directly leads to \eqn{expand}. In this way one has proven a powerful
non-renormalisation theorem which states that perturbatively there is just
the tree-level contribution and the one-loop term $F^{(0)} (X^0,X^A)$.
The Peccei-Quinn symmetry is continuous at the perturbative level. This
indicates that the axion $a$ describes the same physical degrees of freedom
as the dual antisymmetric tensor $\cB_{\mu\nu}$ which is familiar from the
the standard world-sheet formulation of heterotic strings. As usual this
continuous symmetry is broken to its discrete subgroup at the full quantum
level due to space-time instanton effects. These give rise to the
non-perturbative $F^{(k)}(X^0,X^A)\,{\rm exp}[ -2\pi k S]$ terms.

The heterotic string is invariant under a $SO(2,n)$ T-duality group
\cite{GPR}, which is expected to survive at the non-perturbative level
because it can be viewed as a discrete gauge symmetry \cite{DHS}. The
$SO(2,n)$ transformations can be embedded into the symplectic group
$Sp(2(n+2);\Zbar)$ and thus naturally act on the vector multiplets
contained in the low-energy effective action. These $SO(2,n)$
symplectic transformations are most easily written down in a symplectic
basis which is {\it different} from the $STU$ basis. This new basis is
completely specified by the ``stringy" symplectic vector $(\cX^I, \cF_I)$
which is related to the $STU$ symplectic vector $(X^I, F_I
\mathrel{\mathop{=}\limits^{\rm def}} \frac{\d F_{\rm het}(X)}{\d X^I})$ in
the following way \cite{CDFVP}:
\bq \left(\begin{array}{c}  \cX^0 \\ \cX^1 \\
    \cX^A \\ \cF_0 \\ \cF_1 \\  \cF_A
    \end{array}\right) =
    \left(\begin{array}{cccccc} 1&0&0&0&0&0 \\0&0&0&0&1&0 \\
    0&0&{\d^A}_B&0&0&0 \\ 0&0&0&1&0&0 \\ 0&-1&0&0&0&0 \\
    0&0&0&0&0&{\d_A}^B \end{array}\right)
    \left(\begin{array}{c}  X^0 \\ X^1 \\
    X^A \\ F_0 \\ F_1 \\ F_A
    \end{array}\right)
    \, ,
\label{go2stringy}
\eq
In the stringy basis the $SO(2,n)$ transformations are of the
semi-classical type, and are given by
\bq \left(\!\begin{array}{c} \cX^I \\[1mm] \cF_I
     \end{array}\!\right)
     \mathrel{\mathop{\longrightarrow}\limits^{\rm SO(2,n)}}
     \left(\begin{array}{cc} \Uu^I{}_J & 0 \\[1mm]
     (\Uu^{-1})^T{}_I{}^K\, \Lambda_{KJ} &
     (\Uu^{-1})^T{}_I{}^J  \end{array}\right)
     \left(\!\begin{array}{c} \cX^J \\[1mm] \cF_J
     \end{array}\!\right)     \, ,
\label{tdual}
\eq
where
\bq [ \Uu^T \eta \, \Uu ]_{IJ} = \eta_{IJ} \hspace{2cm}
    \Lambda_{IJ} =  \mbox{symmetric, real} \, .
\label{so2n}
\eq
The matrices $\Lambda_{IJ}$ are absent at the string tree-level. They
must be introduced at the one-loop level and also non-perturbatively to
accomodate for the monodromies generated by encircling the singularities in
the quantum moduli space. See e.g. \cite{AFGNT,BHW} for more details
concerning this point.

The relations \eqn{go2stringy} - \eqn{so2n} have some important
consequences. First of all one verifies that
\bq \cF_I = -iS \cX\eta_I + \sum_{k=0}^{\infty} e^{\textstyle -2\pi kS}
    \der_I F^{(k)}(X^0,X^A)
\label{cF}
\eq
where the $\der_I$ stand for partial functional derivatives with respect to
the original scalars $X^I$. Note that we have conveniently included the
the one-loop term ($k=0$) into the instanton sum. In the $I=1$ version of
the equation \eqn{cF} the one-loop and non-perturbative contributions
clearly vanish, so one can derive from the variation of $\cF_1$ how the
dilaton-axion field $S$ behaves under the T-dualities. This yields
\bq S \mathrel{\mathop{\longrightarrow}\limits^{\rm SO(2,n)}}
    S + \sum_{k=0}^{\infty} i
    \frac{\der_I F^{(k)} (\Uu^{-1})^I{}_1 }
    {[\cX\eta\, \Uu^{-1}]_1} \, e^{\textstyle -2\pi k S}
    + i \frac{[\cX\Lambda\, \Uu^{-1}]_1}{[\cX\eta\, \Uu^{-1}]_1} \, .
\eq
In other words, the $N=2$ special coordinate $S$ {\it is only invariant
under $SO(2,n)$ transformations in the classical limit}, when the
$F^{(0)}(X^0,X^A)$, $F^{(k>0)}(X^0,X^A)$ and $\Lambda_{IJ}$ dependent terms
vanish. From the moment on that one-loop and non-perturbative effects are
taken into account $S$ is no longer inert under $SO(2,n)$ transformations,
and even ceases to be single-valued due to the non-trivial monodromies.
Therefore it is best to perform a change of coordinates on the K\"ahler
manifold which amongst others trades the special coordinate $S$ for a
single-valued and $SO(2,n)$ invariant alternative, which we call $S_{\rm
hol}$. This field $S_{\rm hol}$ was introduced in \cite{dWKLL} at the
perturbative level and in \cite{BHW} non-perturbatively, and it turns
out to be a complicated holomorphic function of the moduli $S$ and $Z^A$.
Therefore $S_{\rm hol}$ is not a $N=2$ special coordinate. In the
non-perturbative case one defines
\bq S_{\rm hol} = \frac{i}{(n+2)} \bl \eta^{IJ} \cF_{IJ} + L \br \,.
\label{Shol}
\eq
In order to properly understand this last formula a few more remarks must
be added. Equation \eqn{go2stringy} implies that
\brr &&\cX^I = (X^0, \cX^1, X^A )  \nonumber\\
     &&\cX^1 = -\frac{1}{2} \frac{\eta_{AB} X^A X^B}{X^0} +
     \sum_{k=1}^{\infty} \frac{2\pi i k}{X^0} F^{(k)}(X^0,X^A)\,
     e^{\textstyle -2\pi k S} \, .
\err
As has been emphasized in \cite{BHW} the $\cX^I$ in general define a set of
independent variables. However, this is no longer true in the perturbative
regime, when the instanton terms are supressed. In that case
$\cX^1$ is a function of $(X^0, X^A)$ only, which is reflected in the
constraint
\bq \cX\eta\cX = \sum_{k=1}^{\infty} 4\pi i k\,  F^{(k)}(X^0,X^A)\,
    e^{\textstyle -2\pi k S}
    \mathrel{\mathop{\longrightarrow}\limits^{\rm perturb.}}  0 \, .
\eq
So it is only for finite $S$ that the stringy scalars $\cX^I$ are really
independent, which implies that only in that case one can define a stringy
prepotential and derivatives thereof
\bq \cF(\cX) = \ft{1}{2} \cX^I \cF_I  \hspace{2cm}
    \cF_{IJ} = {\check \der_I} {\check \der_J} \cF  \, .
\label{checkprep}
\eq
In the perturbative case a prepotential doesn't exist, so the
abstract expression \eqn{Shol} doesn't make sense there. Nevertheless one
can work out \eqn{Shol} in terms of the functions $F^{(k)}(X^0,X^A)$ and
then take the perturbative limit. This procedure leads to the perturbative
expression for $S_{\rm hol}$ as it was given in \cite{dWKLL}:
\bq S_{\rm hol} \mathrel{\mathop{\longrightarrow}\limits^{\rm perturb.}}
    S + \frac{i}{(n+2)} \bl \eta^{AB} F_{AB}^{(0)} + L^{(0)} \br  \, .
\label{Sholpert}
\eq
The function $L$ and its semi-classical limit $L^{(0)}$ are necessary to
cancel the infinities contained in $\eta^{IJ} \cF_{IJ}$ and $\eta^{AB}
F_{AB}^{(0)}$ respectively. Moreover, one must impose $L \rightarrow L -
\eta^{IJ} \Lambda_{IJ}$ in order to keep $S_{\rm hol}$ invariant under the
monodromies.

Although $S_{\rm hol}$ clearly is a natural coordinate on the special
K\"ahler manifold, it is still not describing the true heterotic dilaton
$\phi_{\rm inv}$. Perturbatively one finds \cite{GreenSchw} that
\bq \phi_{\rm inv} \mathrel{\mathop{=}\limits^{\rm perturb.}}
    \phi + \frac{i\bX^I F_I^{(0)}(X) + h.c.} { 2 \cX\eta\cbX } \, ,
\label{phiinv}
\eq
where the second term is the so-called Green-Schwarz term.  
At this point one may want to introduce yet another complex scalar, called
$S_{\rm inv}$, which contains the invariant dilaton $\phi_{\rm inv}$ as its
real part. We propose the following definition, which makes sense at the
full non-perturbative level and reduces to \eqn{phiinv} semi-classically:
\bq S_{\rm inv} = i \frac{\cF_I \cbX^I}{\cX\eta\cbX} -iM
    \mathrel{\mathop{=}\limits^{\rm def}} \phi_{\rm inv} -i a_{\rm inv}\, .
\label{Sinv}
\eq
Here $M$ is a real function transforming as
\bq M \rightarrow M + \frac{\cX\Lambda\cbX}{\cX\eta\cbX} \, ,
\eq
which ensures that $S_{\rm inv}$ is monodromy invariant. Note that
$S_{\rm inv}$ is a {\it non-holomorphic} function of the special 
coordinates $S,Z^A$ so one cannot use it as a prefered K\"ahler coordinate.
However, all the physical couplings in the effective action should be 
expanded in terms of the true dilaton-axion complex $S_{\rm inv}$ in order 
to properly identify the perturbative and non-perturbative contributions.

\subsection{The ``stringy" vector fields}

In \eqn{go2stringy} we introduced the stringy symplectic vector $(\cX^I,
\cF_I)$ in terms of which the T-duality transformations acquire a simple
form. Of course one can also write the T-dualities in terms of the $STU$
symplectic vector $(X^I,F_I)$ in which case --- in the notation of
\eqn{basicvec} --- a $Z^{IJ}\neq 0$ term is generated. This means that some
of the field strengths $F_{\mu\nu}^I$ transform into their ``duals"
$G_{\mu\nu I}$ under the T-dualities, which implies that the $STU$ vectors
$W^I_\mu$ themselves transform in a non-local way. This state of affairs
indicates that the $STU$ vectors $W_\mu^I$ are not the most natural objects
to work with. As was explained in \cite{CDFVP} one may proceed by trading
the vector $W_\mu^1$ for a new vector $\cW_\mu^1$ via a duality
transformation. In this way one gets the so-called stringy vector fields
\bq \cW_\mu^I = (W_\mu^0, \cW^1_\mu, W^A_\mu) \, .
\eq
By construction these satisfy
\bq
2 \der_{[\mu} \cW_{\nu] }^I = \cF_{\mu\nu}^I =
(F_{\mu\nu}^0, G_{\mu\nu 1}, F_{\mu\nu}^A ) \, ,
\eq
from which one easily derives that the stringy vectors $\cW^I_\mu$
transform just linearly into eachother under the T-dualities.

Having established that the stringy variables are the most natural ones to
work with, we must still explain how one constructs the stringy version of
the vector multiplet lagrangian. At the non-perturbative level everything
is straightforward because one may effectively implement the duality
transformation on the vector $W_\mu^1$ by applying the symplectic
transformation \eqn{go2stringy}. This leads to the prepotential $\cF(\cX)$
of \eqn{checkprep} which can be inserted into the general superconformal
action \eqn{lagr}. In the perturbative case a sensible stringy prepotential
doesn't exist because the $\cX^I$ are not independent. As a result the
superconformal action formula \eqn{lagr} cannot be used as it stands.

In what follows we will explore in which sense the superconformal approach
fails to deal with the perturbative effective action in the stringy basis.
In particular we will show that one can actually stay pretty close to the
familiar superconformal ideas and still obtain the desired theory. We take
the superconformal action in the $STU$ basis as a starting point and
explicitly dualise the vector $W_\mu^1$. In doing so we will clearly see
which ingredient of superconformal supergravity is incompatible with the
dualisation procedure. The crucial point is that one is forced to eliminate the
auxiliary field $T_{\mu\nu\, ij}^+$ contained in the Weyl multiplet in
course of the dualisation. As a result the Weyl multiplet is no longer
realised off-shell. Below we exhibit the results of the necessary
computations in rather detail, amongst others because it sets the stage for
the other duality transformation we intend to perform, namely the
dualisation of the heterotic axion $a$. Both dualisations share a lot of
common features, as will become clear in section 5.

We consider the lagrangian \eqn{lagr} with as prepotential the perturbative
expression
\bq  F (X^I) = -\frac{1}{2} \frac{X^{1}}{X^0} \eta_{AB} X^A X^B
     + F^{(0)}(X^0,X^A)  .
\label{prep}
\eq
From now on we will always work in the perturbative regime, which from a
supergravity point of view is the most interesting one. The stringy
variables $\cX^I$ and $\cO_i^I$ are then given by
\brr &&\cX^I = (X^0, \cX^{1}, X^A) \nonumber\\
     &&\cO^I_i \,= ( \O^0_i, \cO^{1}_i, \O^A_i) \nonumber\\
     &&\cX^{1} \mathrel{\mathop{=}\limits^{\rm def}}
     F_{1} = -\frac{1}{2} \frac{\eta_{AB} X^A X^B }{X^0}  \nonumber\\
     &&\cO^{1}_i \,\mathrel{\mathop{=}\limits^{\rm def}}
     F_{1I}\O^I_i = - \frac{\cX^{1}}{X^0} \O^0_i- \eta_{AB}
     \frac{ X^A}{X^0} \O^B_i  \, .
\label{checkvar}
\err
The vector $W_\mu^1$ can be dualised by treating the field strength
$F_{\mu\nu}^1$ as an independent variable on which a Bianchi identity
has been imposed by means of a lagrange multiplier $\cW^1_\mu$. This
leads to the lagrangian
\bq e^{-1} \Ll_{\rm vector} + \ft{i}{4} e^{-1} \e^{\mu\nu\l\s}
    F_{\mu\nu}^1\der_\l \cW^1_\s \, .
\label{actioncW}
\eq
For future use we list the $F_{\mu\nu}^1$ dependent terms in the lagrangian:
\brr e^{-1} \Ll_{\rm vector} &=& e^{-1} \Ll_{\rm vector}
     \Big|_{F_{\mu\nu}^1=0} \nonumber\\
     &&-\ft{i}{4} F_{\mu\nu}^{+1} \bpl \bar{F}_{1I} \hF^{\mu\nu +I}
     - \ft{i}{2} XN_{1} T^{\mu\nu +}_{ij} \ve^{ij} -\ft{1}{4} \bar{F}_{1IJ}
     \bar{\O}^{iI}\s^{\mu\nu} \O^{jJ}\ve_{ij} \bpr +h.c.\nonumber\\
     &&+\ft{i}{8} e^{-1} \e^{\mu\nu\l\s} F_{\mu\nu}^1
     \bpl \cbO{}^{1}_i \g_\l \psi_{\s j}\ve^{ij} -\cX^{1} \bpsi_{\l i}
     \psi_{\s j} \ve^{ij} \bpr + h.c.
\label{F1terms}
\err
Note that \eqn{F1terms} depends at most linearly on $F_{\mu\nu}^1$. This
is a direct consequence of the fact that the prepotential \eqn{prep} is at
most linear in $X^1$. The action \eqn{actioncW} ---which still contains all
the auxiliary fields of the Weyl multiplet--- is most suitable for
determining the supersymmetry transformation of the Lagrange multiplier
$\cW^1_\mu$. The supersymmetry transformation rules for the original $STU$ 
vector multiplet
fields are as in \eqn{vrules}, except for $\d W_\mu^1$ which is replaced by
\bq \d F_{\mu\nu}^1 = 2 \der_{[\mu} \bl \bar{\e}_i\g_{\nu ]} \O_j^{\,1}
    \ve^{ij} +2\bX^1  \bar{\e}^i\psi_{\nu ]}^j \ve_{ij} +h.c. \br \, .
\eq
The variation of the $\cW^1_\mu$-independent part of the lagrangian is
necessarily of the following form
\bq \d\, \Ll_{\rm vector} =  -\ft{i}{4} \e^{\mu\nu\l\s}
    {\cal O}_\mu\, \der_\nu  F_{\l\s}^1 + \mbox{total derivatives}
\, ,
\eq
otherwise one would not regain invariance if the Lagrange multiplier would
be eliminated again. Here ${\cal O}_\mu$ stands for an a priori unknown
expression whose precise form can be determined by direct computation.
Transforming $F_{\mu\nu}^1$ in the Lagrange multiplier term of
\eqn{actioncW} generates a total derivative, so if one lets the variation
of $\cW_\mu^1$ be equal to ${\cal O}_\mu$ one obtains invariance. This
reasoning yields
\bq \d \cW_\mu^1 = \bar{\e}_i\g_\mu \cO_j^{\,1} \ve^{ij}
    +2\cbX{}^1  \bar{\e}^i\psi_\mu^j \ve_{ij} +h.c. \, ,
\label{delW1}
\eq
Note that $\d W_\mu^0, \d W_\mu^A$ and $\d \cW_\mu^1$ nicely rotate into
eachother under $SO(2,n)$ transformations, as expected.

In order to finish
the dualisation we eliminate the auxiliary field $F_{\mu\nu}^1$. From
\eqn{F1terms} one derives that the equation of motion enforced by
$F_{\mu\nu}^1$ reads
\bq 4\cbX\eta_I \hcF{}_{\mu\nu}^{+I} - \cX\eta\cbX\,  T_{\mu\nu \, ij}^+
    \ve^{ij} - \eta_{IJ} \cbO{}^{iI}\s_{\mu\nu} \cO^{jJ}\ve_{ij} = 0 \, ,
\label{F1eom}
\eq
where the covariant field strength $\hcF{}_{\mu\nu}^{+1}$ is defined in the
same spirit as in \eqn{covstrength}. Remark that \eqn{F1eom} is
saturated at the string tree level in the sense that it doesn't
receive any loop corrections. Moreover it is manifestly $SO(2,n)$
invariant. However, the most important property of the $F_{\mu\nu}^1$
equation of motion is that it doesn't determine the value of
$F_{\mu\nu}^1$ itself,
at least as long as one treats $T_{\mu\nu \, ij}^+$ as an independent
auxiliary field. But in fact $T_{\mu\nu \, ij}^+$ cannot be kept as an
independent variable, because \eqn{F1eom} must be viewed as a constraint
determining $T_{\mu\nu \, ij}^+$ as a function of the stringy field
strengths, scalars and gauginos\footnote{The factor $\cX\eta\cbX$ can be
inverted, otherwise the $R$ term in the action would vanish.}.
As a result the $T_{\mu\nu \, ij}^+$ equation of motion \eqn{Teomgen}
must be imposed as well. In the present case this equation of motion
can be worked out, yielding
\brr &&\cZ\eta\cbZ \bpl F_{\mu\nu}^{+1} + i \bar{S} F_{\mu\nu}^{+0}
     -i\phi X^0 T_{\mu\nu \, ij}^+ \ve_{ij} \bpr \nonumber\\
     &&= 4 \phi\,\cF_{\mu\nu}^{+I}\eta_{IJ}\, {\rm Re}\,\cZ^J +
     \mbox{fermions}^2 + F^{(0)}\mbox{terms}
     + \mbox{hypermultiplet terms}\, ,
\label{exprF1}
\err
where $\cZ^I = -i\frac{\cX^I}{X^0}$. In this way $F_{\mu\nu}^1$
finally becomes a dependent expression of the remaining physical fields.
Remark that the actual expression for $F_{\mu\nu}^1$ receives loop and
hypermultiplet dependent contributions.

To finish our construction of the vector multiplet theory in the stringy
basis we discuss its symplectic properties. We start from the lagrangian
\eqn{actioncW}, impose the two equations of motion we just mentioned and
call the resulting lagrangian $e^{-1} \Ll_{\rm stringy}$. In fact the
expression \eqn{F1eom} is nothing but  a nice way of writing that
$\cF^{\mu\nu\, +1} =  G_1{}^{\mu\nu\, +}$. Moreover we define that
$\cG_1{}^{\mu\nu\, +} = -F^{\mu\nu\, +1}$. The symplectic transformations
of $\cF^{\mu\nu\, +1}$ and $\cG_1{}^{\mu\nu\, +}$ are by definition given
by the transformations of $G_1^{\mu\nu\, +}$ and $-F^{\mu\nu\, +1}$
respectively\footnote{Notice that $F_{\mu\nu}^{+1}$ as a dependent field
still transforms in the naive way due to the fact that the $T_{\mu\nu \,
ij}^+ $ equation of motion is symplectically invariant.}. One may
verify that with these definitions the stringy action is symplectically
invariant apart from the non-invariant term
\bq -\ft{i}{8} F_{\mu\nu}^{+I} G^{\mu\nu\, +}_I
    +\ft{i}{4} F_{\mu\nu}^{+1} \cF^{\mu\nu\, +1} +h.c.
    = -\ft{i}{8} \cF_{\mu\nu}^{+I} \cG^{\mu\nu\, +}_I +h.c. \, .
\label{stringysympl}
\eq
In particular this implies that the stringy lagrangian transforms as
follows under $SO(2,n)$ transformations:
\bq e^{-1} \Ll_{\rm stringy} \mathrel{\mathop{\longrightarrow}\limits^{\rm
    SO(2,n)}} e^{-1} \Ll_{\rm stringy} - \ft{i}{8} \Lambda_{IJ}
    \cF_{\mu\nu}^I   \,\tilde{\!\cF}{}^{\mu\nu\, J} \, .
\label{sotransfoaction}
\eq

\setcounter{equation}{0}
\section{Construction of generic antisymmetric tensor theories}

In this section we take a general point of view and study arbitrary vector
multiplet theories containing a Peccei-Quinn symmetry. We impose just one
restriction on this class of theories, namely we demand that there exists a
set of special coordinates on which the Peccei-Quinn transformation acts in
a standard way. By this we mean that there exists one distinguished
special coordinate, henceforth called $S$, which shifts under the
Peccei-Quinn symmetry by an imaginary constant, whereas the other special
coordinates $Z^A$ are invariant under it\footnote{The fields $S$, $Z^A$ are
the obvious generalisations of the variables we encounter in the
description of perturbative heterotic strings. The interpretation of $S$ as
a string coupling constant is of course not valid in the general case.}.
Given this ansatz we are able to completely characterise the Peccei-Quinn
invariant vector multiplet theories and we show that they precisely
comprise the cases discussed in \cite{wij} plus the perturbative heterotic
string case which was still missing there.

For all these theories we can obtain an antisymmetric tensor 
version by dualising the axion $a = - {\rm Im} S$. This dualisation can be
performed in a way which is to a large extent model independent and as such
it explains why the resulting antisymmetric tensor theories show some 
universal behaviour. We will see for instance that they are all
characterised by a similar gauge structure. They contain a particular
$U(1)$ gauge symmetry, with parameter $z$, under which the antisymmetric 
tensor field $B_{\mu\nu}$ and an appropriately defined vector gauge field  
$V_\mu$ transform in the following way:
\brr V_\mu &\rightarrow& z V_\mu^{(z)}\nonumber\\
     B_{\mu\nu} &\rightarrow& z B_{\mu\nu}^{(z)}
\, .
\err
Here the fields $V_\mu{}^{(z)}$ and $B_{\mu\nu}{}^{(z)}$ stand for some
(complicated) functions of the independent fields in our superconformal
theory which will be specified below. In the treatment of \cite{wij} this 
$z$ gauge symmetry coincides with the central charge transformation which 
is necessary for off-shell closure of the supersymmetry algebra.

At a later stage we verify whether or not one may dualise the vector
$V_\mu$ and we find that the different antisymmetric tensor theories
behave differently in this respect. The so-called non-linear 
vector-tensor multiplet theory of \cite{wij} doesn't allow for this 
dualisation. The linear vector-tensor theory \cite{wij} can be 
dualised. This last theory is known to contain two 
abelian background vector fields, one of which gauges the central charge 
transformation. Under the duality the role of both background vectors gets
interchanged. In the heterotic case, finally, the dualisation
removes the complete central charge-like structure. We will see later that
in this last case the dual of $V_\mu$ is nothing but the stringy vector
$\cW_\mu^1$ we encountered before. So the stringy vectors play at least two
important roles. We saw in section 3.2 that they make the $SO(2,n)$
invariance of the heterotic vector multiplet theory as manifest as possible,
and at present we find that they also considerably simplify the gauge
algebra acting on the heterotic vector and tensor gauge fields.

\subsection{Peccei-Quinn invariant vector multiplet theories}

The aim of this subsection is to characterise the set of Peccei-Quinn
invariant $N=2$ vector multiplet theories. We demand that any theory in
this set contains some special coordinate moduli $(S,Z^A)$ transforming in
the following way under a continuous Peccei-Quinn symmetry:
\brr S &\mathrel{\mathop{=}\limits^{\rm def}}& -i \frac{X^1}{X^0}\,
     \mathrel{\mathop{\longrightarrow}\limits^{\rm P.Q.}}  S -i c
     \nonumber\\
     Z^A &\mathrel{\mathop{=}\limits^{\rm def}}& -i \frac{X^A}{X^0}
     \mathrel{\mathop{\longrightarrow}\limits^{\rm P.Q.}} Z^A \, ,
\label{PQmod}
\err
Here $c$ is an arbitrary real constant. In order to understand how the
Peccei-Quinn symmetry acts on a the full vector multiplet theory (so not
only on the scalars) we embed the Peccei-Quinn transformation into the
symplectic group $Sp(2(n+2),\Rbar)$
\bq
\left(\begin{array}{cc}  {U^I}_J &Z^{IJ} \\[1mm]  W_{IJ} &{V_I}^J
\end{array}\right)_{\rm P.Q.}
= \left(\begin{array}{cccccc} 1&0&0&0&0&0 \\c&1&0&0&0&0 \\
0&0&{\d^A}_B&0&0&0 \\
W_{00} &-\frac{c}{2}W_{11}+ W_{1} &-\frac{c}{2}W_{1B}+ W_{B} &1&-c&0 \\
\frac{c}{2}W_{11}+ W_{1} &W_{11} & W_{1B} &0&1&0 \\
\frac{c}{2}W_{1A}+ W_{A} &W_{1A} & W_{AB}&0&0&{\d_A}^B
\end{array}\right) \, .
\label{PQ}
\eq
This particular symplectic transformation is a crucial object in the
construction of antisymmetric tensor theories, so we will comment on some
of its characteristic features. First we note that $Z^{IJ}=0$, so the
symplectic transformation is of the semi-classical type. In particular
this means that the Peccei-Quinn symmetry transforms the vectors $W^I_\mu$
just linearly into eachother.
Note that the submatrices ${U^I}_J$ and ${V_I}^J = (U^{-1})^T{}_I{}^J$ are
completely fixed by the transformations \eqn{PQmod} of the special
coordinate moduli and by the defining symplectic relation \eqn{defsympl},
so they are valid for all cases we are interested in.
The submatrix $W_{IJ}$ is model dependent. In equation \eqn{PQ} the most
general matrix $W_{IJ}$ has been given which is consistent with
\eqn{defsympl} (provided $W_{AB}$ is taken to be symmetric), but
in fact there are additional restrictions. In any case the various entries
of $W_{IJ}$ must be real constants. They may depend on the parameter $c$
but not for instance on any of the moduli fields. The additional
restrictions we just mentioned stem from the fact that the symplectic
transformation \eqn{PQ} must generate a {\it duality invariance} of the
theory and not just a symplectic reparametrisation. This implies that
\bq
F_I(X^0,X^1,X^A) \mathrel{\mathop{\longrightarrow}\limits^{\rm P.Q.}}
F_I(X^0,X^1 + cX^0,X^A) \, ,
\eq
so when a suitable prepotential $F(X)$ is given the explicit form of
$W_{IJ}$ can be read off easily. Nicely enough the reverse procedure is
possible as well. As we are going to explain now, one can selfconsistently
solve for the most general matrix $W_{IJ}$ leading to a duality invariance.
The most general prepotential $F(X)$ can then be reconstructed from the
knowledge of $W_{IJ}$.

We recall that the lagrangian \eqn{lagr} is Peccei-Quinn invariant up to a
topological term
\bq e^{-1} \Ll_{\rm vector} \mathrel{\mathop{\longrightarrow}
    \limits^{\rm P.Q.}} e^{-1} \Ll_{\rm vector}
    -  \ft{i}{8} (U^T W)_{IJ} F_{\mu\nu}^I {\tilde F}^{\mu\nu \,  J}\, ,
\label{noninv}
\eq
thanks to the fact that the transformation \eqn{PQ} is semi-classical. In
terms of the standard vector multiplet variables the non-invariance of the
lagrangian may have different sources, because not only the axion $a$
shifts under Peccei-Quinn transformations, but also $W_\mu^1$, $\O^1_i$ and
$Y^1_{ij}$ transform due to the equation \eqn{PQ}. The situation becomes
much more transparant though, if one performs a change of variables and
rather works with a set of variables which --- apart from the axion
itself--- are {\it all} invariant under Peccei-Quinn. Given equation
(\ref{PQ})
one easily verifies that the following variables do the job:
\brr &&a\, , \phi= {\rm Re} \,S\, , X^0\, , X^A  \nonumber\\
     &&\l_i \mathrel{\mathop{=}\limits^{\rm def}}
     {\displaystyle \frac{-i}{2X^0}} \bl \O^{1}_i -
     {\displaystyle \frac{X^{1}}{X^0}} \O^0_i \br\, ,  \O^0_i\, ,  \O^A_i
     \nonumber\\
     && V_\mu \mathrel{\mathop{=}\limits^{\rm def}} W_\mu^{1} - a W_\mu^0
     \, , W^0_\mu\, , W^A_\mu \nonumber\\
     && Z_{ij} \mathrel{\mathop{=}\limits^{\rm def}} Y_{ij}^{1} - a
     Y_{ij}^0\, , Y_{ij}^0 \, , Y_{ij}^A
\label{newvar}
\err
When the lagrangian is expressed in terms of \eqn{newvar} there is only
one source for the non-invariance term in \eqn{noninv}: it must come
from terms in the lagrangian in which the axion $a$ appears
undifferentiated. As an example we concentrate on the term
\bq - \ft{i}{8} (U^T W)_{IJ} F_{\mu\nu}^I {\tilde F}^{\mu\nu \,  J}
    = -  \ft{i}{8} W_{AB}(c) F_{\mu\nu}^A {\tilde F}^{\mu\nu \,  B}
    + \cdots
\eq
This term should arise upon varying
\bq e^{-1} \Ll_{\rm vector} = -\ft{i}{8} {\rm Re} F_{AB} \,
    F_{\mu\nu}^A\, {\tilde F}^{\mu\nu \,  B} + \cdots
\eq
from which we derive that the following relation must be valid:
\bq {\rm Re}F_{AB}(a \mapsto a+c) = {\rm Re}F_{AB}(a) + W_{AB}(c)
\eq
It is then clear that
\brr W_{AB}(c) &=& -c\, \eta_{AB} \nonumber\\
    {\rm Re}F_{AB}(a) &=& - a\,\eta_{AB}+\mbox{$a$ independent terms} \, ,
\err
where $\eta_{AB}$ is a real symmetric constant {\it not} depending on the
parameter $c$. In the same spirit one can determine the other entries
in $W_{IJ}$ by writing \eqn{noninv} in terms of the Peccei-Quinn
invariant field strengths $F_{\mu\nu}^0, 2 \der_{[\mu} V_{\nu ]},
F_{\mu\nu}^A$ and by reconstructing the undifferentiated $a$ terms in the
lagrangian which lead to \eqn{noninv}. The result is
\bq W_{IJ} = -c \left(\begin{array}{ccc}
    \eta_{00} - \frac{c^2}{6}\eta_{11} &\eta_{01} - \frac{c}{2}\eta_{11}
    &\eta_{0B} - \frac{c}{4}\eta_{1B} \\
    \eta_{01} + \frac{c}{2}\eta_{11}&\eta_{11}&\frac{1}{2}\eta_{1B}\\
    \eta_{0A} + \frac{c}{4}\eta_{1A}&\frac{1}{2}\eta_{1A}&\eta_{AB}
    \end{array}\right) \, ,
\label{Wexplit}
\eq
where all $\eta_{IJ}$ are constants.
This result can be integrated back to yield the prepotential\footnote{In
order to make contact to the perturbative heterotic prepotential given
in \eqn{prep} all entries of $\eta_{IJ}$ apart from $\eta_{AB}$
must be put equal to zero. Therefore the $\eta_{IJ}$ which is used throughout
this section doesn't coincide with the $SO(2,n)$ metric defined in
\eqn{SOmetric}. Note that all $\eta_{IJ}$'s occuring outside section 4 
refer to the $SO(2,n)$ metric and not to the matrix $\eta_{IJ}$
we just defined.  }
\brr F(X) &=& -\frac{1}{6}\frac{(X^1)^3}{X^0}\eta_{11}
    -\frac{1}{4} \frac{(X^1)^2}{X^0} \eta_{1A} X^A
    -\frac{1}{2} \frac{X^1}{X^0} \eta_{AB} X^A X^B + F^{(0)}(X^0,X^A) \nonumber\\
    &&- \frac{1}{2} \bl(X^1)^2 \eta_{01} + X^1 X^0\eta_{00}+2X^1 X^A
    \eta_{0A}\br
\label{genprep}
\err
Note that the $\eta_{0I}$ terms in the prepotential are quadratic
polynomials with real coefficients. These just add total divergence
terms to the action \cite{dWLVP} so from now on we put $\eta_{0I}=0$ for
simplicity. As such we precisely obtain the theories discussed in
\cite{wij} and on top of that the perturbative heterotic string case
$\eta_{11}=0$, $\eta_{1A}=0$ which was excluded in a fully off-shell
superconformal context. We see no sign of any other vector multiplet 
theories which might have a dual antisymmetric tensor description. Remark 
that the prepotential \eqn{genprep} has a natural type II interpretation.
For type II strings on generic Calabi-Yau manifolds there is a 
Peccei-Quinn symmmetry for every modulus $Z^\Aa$ \cite{Cremmer,dWVVP}. One
may pick any value for $\Aa$
and identify the corresponding modulus with $S$. In the large $S$ limit the
prepotential \eqn{prepII} coincides with the prepotential we just found,
provided we take $d_{111} = \eta_{11}$, $d_{11A}= \ft{1}{2} \eta_{1A}$, 
$d_{1AB}= \eta_{AB}$.

\subsection{Dualising the axion}

All the Peccei-Quinn invariant theories we just specified can be
dualised into antisymmetric tensor theories. In order to check that this 
is indeed possible, it suffices to show that the vector multiplet lagrangians
for these theories can be brought into a form such that the axion $a$
appears only via its ``field strength" $\der_\mu a$ (up to total derivative
terms in the lagrangian). The existence of the variables \eqn{newvar} is
very important in this respect. When the lagrangian \eqn{lagr} is written
in terms of the standard vector multiplet variables one sees
undifferentiated $a$ dependencies occurring at various places. By going to
the new variables one effectively absorbs most of these unwanted $a$ terms.
Only in the gauge sector some undifferentiated $a$ terms remain. As we
remarked before these left-over $a$ terms can be reconstructed on the base
of equation \eqn{noninv}. They read
\brr
&&\ft{i}{8} \left\{ a\, \left( \eta_{11} F_{\mu\nu}(V) \tilde
F^{\mu\nu}(V)
+ \eta_{1A} F_{\mu\nu}(V) \tilde F^{\mu\nu A} + \eta_{AB} F^A_{\mu\nu}
\tilde F^{\mu\nu B} \right) \right. \nonumber\\
&& \left.\phantom{\ft{i}{8} \{ }+ a^2 \, \left( \eta_{11} F_{\mu\nu}(V)
\tilde F^{\mu\nu 0}
+ \ft{1}{2} \eta_{1A} F_{\mu\nu}^0 \tilde F^{\mu\nu A} \right)
+ \ft{1}{3} a^3\eta_{11} F_{\mu\nu}^0 \tilde F^{\mu\nu 0} \right\} \, ,
\label{expla}
\err
where the gauge covariant field strength $F_{\mu\nu}(V)$ is defined as
\bq
F_{\mu\nu}(V) = 2 \der_{[\mu } V_{\nu ]}
- 2 W_{[\mu }^0 \der_{\nu ]}^{\phantom 0} a \,.
\eq
Equation \eqn{expla} can be rewritten as
\bq - \ft{i}{4}  e^{-1} \e^{\mu\nu\l\s} \der_\mu a\,
    \bpl \eta_{11} V_\nu  \der_\l V_\s + \eta_{1A} W_\nu^A  \der_\l V_\s
    + \eta_{AB} W_\nu^A  \der_\l W_\s^B \bpr + e^{-1} \bpl \mbox{total
    derivative} \bpr \, .
\label{topterm}
\eq
After dropping the total derivative we may dualise the axion by replacing
the ``field strength" $\der_\mu a$ everywhere in the lagrangian by an
auxiliary vector $V_\mu{}^{(z)}$ and by adding a Lagrange multiplier term
\bq \ft{i}{4} e^{-1} \e^{\mu\nu\l\s} V_\mu^{(z)} \, \der_\nu B_{\l\s} \,
. \label{B}
\eq
In principle the auxiliary field  $V_\mu{}^{(z)}$ can then be eliminated
(possibly together with some other auxiliaries) in order to obtain an
antisymmetric tensor theory. This turns  $V_\mu{}^{(z)}$ into a dependent
expression of the physical fields. In section 5.3 we will explicitly
discuss the elimination of  $V_\mu{}^{(z)}$ in the perturbative heterotic
context.

\subsection{The gauge structure of the antisymmetric tensor theories}

It is interesting to see how a central-charge-like gauge structure, which
is a crucial ingredient in the off-shell construction of \cite{wij}, arises
in the present context. The gauge transformations of the vector fields are
easily determined starting from $\d_{\rm gauge} W^I_\mu = \der_\mu \t^I$
and the redefinition \eqn{newvar}. Defining $z = \t^0$, $y= \t^{1} - a \t^0$
one gets:
\bq
\d_{\rm gauge} V_\mu = \der_\mu y + z V_\mu^{(z)} \hspace{1cm}
\d_{\rm gauge} W^0_\mu = \der_\mu z \hspace{1cm}
\d_{\rm gauge} W^A_\mu = \der_\mu \t^A \,.
\eq
Notice that the $z$-gauge transformation maps the vector $V_\mu$ into
$V_\mu{}^{(z)}$ which explains why $V_\mu{}^{(z)}$ appears in the gauge
covariant field strength $F_{\mu\nu}(V)$. The vector field $W^0_\mu$ is a
distinguished one in that it gauges this $z$-gauge transformation. The
gauge transformations of the antisymmetric tensor field $B_{\mu\nu}$
can most easily be determined from the lagrangian
\bq e^{-1} \Ll_{\rm vector} + \ft{i}{4} e^{-1} \e^{\mu\nu\l\s} V_\mu^{(z)}
    \, \der_\nu B_{\l\s} \, ,
\eq
in which $V^{(z)}_\mu$ and the other auxiliaries are kept as independent
non-propagating degrees of freedom. The $B_{\mu\nu}$ independent part of
this lagrangian is not invariant under gauge transformations. First there
are the $F_{\mu\nu}(V)$ dependent terms. Due to the fact that $\d_{\rm
gauge} F_{\mu\nu}(V) = 2 z \der_{[\mu } V_{\nu ]}{}^{(z)}$ they give a
non-zero contribution. Secondly there is the non-total-derivative part of
\eqn{topterm} (with $\der_\mu a$ replaced by $V_\mu{}^{(z)}$) which
contains explicit gauge fields. Note that all these non-invariances are
proportional to $\der_{[\mu } V_{\nu ]}{}^{(z)}$, so they can be cancelled
by a suitable choice of the gauge variation of the antisymmetric tensor
field. In this way one finds that
\brr \d_{\rm gauge} B_{\mu\nu}  &=& 2\der_{[ \mu} \Lambda_{\nu ]} +
     \eta_{11}\,y \, \der_{[\mu } V_{\nu ]}
     + \eta_{1A}\t^A  \der_{[\mu } V_{\nu ]}
     + \eta_{AB}\t^A  \der_{[\mu } W_{\nu ]}^B
     + z\,  B_{\mu\nu}^{(z)}\nonumber\\
     B_{\mu\nu}^{(z)} &=& 4i \tilde{A}_{\mu\nu}
     - \eta_{11} V_{[\mu } V_{\nu ]}^{(z)}
     - \eta_{1A} W_{[\mu }^A V_{\nu ]}^{(z)} \nonumber\\
     A_{\mu\nu} &=& \frac{\textstyle \d \, e^{-1} \Ll_{\rm vector}}
     {\textstyle\d F^{\mu\nu 1}}\Big|_{F_{\mu\nu}^1 \mapsto
     F_{\mu\nu}(V),\, F_{IJ}\mapsto F_{IJ}(a=0)} \, ,
\label{delBgen}
\err
where $\Lambda_\mu$ is a parameter for tensor gauge transformations. Under
the $z$-gauge transformation the antisymmetric tensor is mapped into
$B_{\mu\nu}{}^{(z)}$. This is a complicated function which does not only
contain the field strengths but also scalars, gauginos, gravitinos etcetera.
The gauge covariant field strength for $B_{\mu\nu}$ reads
\brr
H_{\mu\nu\l } &=& \der_{[ \mu} B_{\nu\l ]}
- \eta_{11}\,V_{[ \mu}  \der_\nu V_{\l ]}
- \eta_{1A} W^A_{[ \mu}  \der_\nu V_{\l ]}
- \eta_{AB} W^A_{[ \mu}  \der_\nu W_{\l ]}^B
- W^0_{[ \mu} B^{(z)}_{\nu\l ]} \, .
\err
The constants $\eta_{IJ}$ which were previously introduced in order to
specify the prepotential \eqn{genprep} turn out to be directly related
to the various Chern-Simons couplings of the antisymmetric tensor.
It was already remarked in \cite{wij} that a cubic in $S$ prepotential
leads to a Chern-Simons coupling which is quadratic in the vector
$V_\mu$ and that the quadratic in $S$ prepotential leads to a Chern-Simons
coupling which is linear in $V_\mu$. Here we see that the Chern-Simons 
couplings depend on the background vector multiplet fields only, when the 
prepotential is just linear in $S$.

There is a last point concerning the gauge structure of the antisymmetric 
tensor theories which deserves to be investigated. When we specialise to 
the heterotic string case we see that we have constructed an antisymmetric 
tensor theory containing the vector $V_\mu$ which is directly related to 
the $STU$ vector $W_\mu^1$. We already discussed in section 3.2 that in the
vector multiplet version of the perturbative heterotic theory one may
benefit from trading $W_\mu^1$ for its dual, the stringy vector
$\cW_\mu^1$. One may ask a similar question in the present context,
and verify what are the effects of dualising the vector $V_\mu$. First of
all we note that we have to exclude the $\eta_{11}\neq 0$ case, because
otherwise the theory contains explicit $V_\mu$ terms (see for instance 
\eqn{topterm}), which prevent us from performing the dualisation we have in
mind. In the other cases there is no obstruction, so one may replace the
field strength $2\der_{[\mu} V_{\nu ]}$ by an auxiliary field $C_{\mu\nu}$
and add the Lagrange multiplier term
\bq
\ft{i}{4} e^{-1} \e^{\mu\nu\l\s} C_{\mu\nu} \der_\l V^d_\s
\label{dualC}
\eq
The gauge transformation of $V_\mu^d$ can be fixed in the standard way by
adopting $\d_{\rm gauge} C_{\mu\nu} = 2\der_{[\mu }\{ z V_{\nu
]}{}^{(z)}\}$ and checking the gauge invariance of the theory. One finds
that
\bq
\d_{\rm gauge} V_\mu^d = \der_\mu y^d - \ft{1}{2} \eta_{1A} \t^A 
V_\mu^{(z)} \, .
\label{gaugecV}
\eq
But now we have a extra possibility which crucially hinges
on the fact that $C_{\mu\nu}$ is not a total derivative. We can introduce
an extra gauge transformation $\d_{\rm extra} C_{\mu\nu} = - 2 z \der_{[\mu }
V_{\nu ]}{}^{(z)}$ and still maintain the gauge invariance of the full
theory provided we add a compensating $\d_{\rm extra} B_{\mu\nu}$. This
yields a modified
\bq \d_{\rm gauge} B_{\mu\nu}  = 2\der_{[ \mu} \Lambda_{\nu ]}
    +2 z \der_{[\mu } V_{\nu ]}^d + \eta_{AB}\t^A  \der_{[\mu }
    W_{\nu ]}^B + \ft{1}{2} \eta_{1A}\t^A  C_{\mu\nu}
\label{dualgauget}
\eq
What have we gained by this whole operation? First we look at
the $\eta_{1A}\neq 0$ case. The vector $V_\mu^d$ transforms
under a central charge-like transformation gauged by the vector 
$-\ft{1}{2}\eta_{1A} W_\mu^A$. In going from \eqn{delBgen} to 
\eqn{dualgauget} the complicated
expression $B_{\mu\nu}{}^{(z)}$ has been replaced by the field strength
$2\der_{[\mu } V_{\nu ]}^d$ and conversely the field strength
$2\der_{[\mu } V_{\nu ]}$ has been replaced by $C_{\mu\nu}$. After imposing
the equation of motion for $C_{\mu\nu}$, this field becomes a dependent
expression of the physical fields and as such it can be viewed as a dual
$B_{\mu\nu}{}^{(\eta_{1A} \t^A)}$. So under the duality
transformation the theory has been mapped onto a similar one in which the
role of the vectors $W_\mu^0$ and  $-\ft{1}{2}\eta_{1A} W_\mu^A$ has been
interchanged\footnote{As an example we may take $A=2$, $\eta_{12}=2$. The
vector-tensor theory is then really
mapped onto itself under the duality. This can be checked by going back to
vector multiplet langauge and verifying that the symplectic transformation
\bq
\left(\begin{array}{cc}  {U^I}_J &Z^{IJ} \\[1mm]  W_{IJ} &{V_I}^J
\end{array}\right)_{\rm dual}
= \left(\begin{array}{cccccc} 0&0&1&0&0&0 \\0&0&0&0&1&0 \\
-1&0&0&0&0&0 \\0&0&0&0&0&1 \\0&-1&0&0&0&0 \\ 0&0&0&-1&0&0
\end{array}\right) \, ,
\eq
corresponds to a duality {\it invariance}.}.
In the $\eta_{1A} = 0$ case the whole central charge-like structure
disappears and the Chern-Simons couplings are of a completely conventional
form. We will see later that in that case $V_\mu^d$ can be identified with
the stringy vector $\cW_\mu^1$. The vector fields $W_\mu^0$, $V_\mu^d$ ,
$W_\mu^A$ then appear on an equal footing, reflecting the underlying
$SO(2,n)$ invariance of the heterotic string theory. The resulting
antisymmetric tensor theory will be presented in the next section.

\setcounter{equation}{0}
\section{The antisymmetric tensor effective action for  
         heterotic \newline strings}

\subsection{The $SO(2,n)$ invariant antisymmetric tensor theory}

The results of the preceding section for the particular case of
perturbative strings can be summarised as follows. One starts from the
vector multiplet lagrangian \eqn{lagr} in the $STU$ basis and introduces
the new Peccei-Quinn invariant variables \eqn{newvar}. Then one subtracts
the total derivative term of \eqn{topterm} and adds the $B_{\mu\nu}$
and $V_\mu^d$ Lagrange multiplier terms given in \eqn{B} and \eqn{dualC}
respectively. The order in which the various dualisations are carried out
should not matter so one can equally start from the stringy
vector multiplet theory described in section 3.2 in which the vector
$W^1_\mu$ has already been traded for its dual $\cW_\mu^1$ and afterwards
dualise the axion. Let us sketch the different steps one takes in this
second scenario and verify that it indeed leads to the same theory as in
section 4.3. The advantage of first dualising the vector $W^1_\mu$ is that
one can more easily keep track of the $SO(2,n)$ invariance of the final
antisymmetric tensor theory.

Our starting point is the lagrangian \eqn{actioncW}, or rather $e^{-1}
\Ll_{\rm stringy}$ which is obtained from it by eliminating the auxiliaries
$F_{\mu\nu}^{+1}$ and $T_{\mu\nu \, ij}^+ $. This lagrangian transforms in
a simple way under $SO(2,n)$ transformations, see equation
\eqn{sotransfoaction}. As before one introduces the Peccei-Quinn invariant
variables $\phi, \l_i, Z_{ij}$ given in \eqn{newvar}, as well as a new
field $C'_{\mu\nu}$ which is defined as the Peccei-Quinn invariant part
contained in the dependent field $F_{\mu\nu}^1$:
\bq C'_{\mu\nu}  \mathrel{\mathop{=}\limits^{\rm def}}  F_{\mu\nu}^1 - a
    F_{\mu\nu}^0 \, .
\eq
The stringy vectors $\cW_\mu^I$ are automatically Peccei-Quinn invariant as
can be verified by rewriting the symplectic transformation \eqn{PQ} in the
stringy basis. Therefore the stringy vectors need not be redefined and as
a result no central charge-like gauge transformations appear. The
Peccei-Quinn argument can be repeated, mutatis mutandis, in order to
compute the explicit axion dependence of the lagrangian. One finds that the
only undifferentiated axion term reads
\bq \ft{i}{8}  a_{\rm inv} \,\eta_{IJ} \cF^I_{\mu\nu}
    \, \tilde{\!\cF}{}^{\mu\nu J}
\eq
which can be rewritten as
\bq - \ft{i}{4}  e^{-1} \e^{\mu\nu\l\s} \der_\mu a_{\rm inv} \,
    \eta_{IJ} \cW_\nu^I  \der_\l \cW_\s^J  + e^{-1} \mbox{total
    derivative}  \, .
\label{split}
\eq
Note that in these formulae we introduced the $SO(2,n)$ invariant axion
$a_{\rm inv}$ of \eqn{Sinv} which is related to $a$ by
\bq a_{\rm inv} = a - \frac{1}{2} \frac{ F_I^{(0)}\bX^I+ h.c.
    }{\cX\eta\cbX} + M^{(0)}
\eq
The use of $a_{\rm inv}$ is just a matter of convenience. It guarantees
that the total derivative term in \eqn{split} is $SO(2,n)$ invariant. At
this point we drop the total derivative, replace $\der_\mu a_{\rm inv}$ by
$V_\mu^{(z)}{}_{\rm inv}$ and add the $SO(2,n)$ {\it invariant} Lagrange
multiplier $\cB_{\mu\nu}$. This leads to our final antisymmetric tensor
lagrangian
\bq e^{-1} \Ll_{\rm tensor} = e^{-1} \Big\{ \Ll_{\rm stringy} -
    \mbox{total derivative of \eqn{split}} +
    \ft{i}{4}  \e^{\mu\nu\l\s} V_{\mu \,{\rm inv}}^{(z)} \,
    \der_\nu \cB_{\l\s}\Big\} \, .
\label{vtlagr}
\eq
This lagrangian clearly transforms under the T-dualities as
\bq e^{-1} \Ll_{\rm tensor} \mathrel{\mathop{\longrightarrow}\limits^{\rm
    SO(2,n)}} e^{-1} \Ll_{\rm tensor} - \ft{i}{8}  \Lambda_{IJ}
    \cF_{\mu\nu}^I  \,\tilde{\!\cF}{}^{\mu\nu\, J} \, .
\label{respons}
\eq
One can verify that the lagrangian we just specified equals the one
discussed in section 4.3 provided we identify\footnote{To be completely
precise: both lagrangians differ at the one-loop level by the gauge
invariant total derivative term
\bq  -\frac{i}{4} e^{-1} \e^{\mu\nu\l\s} \der_\mu \Big\{ \Big[ \frac{1}{2}
     \frac{F_I^{(0)} \bX^I + h.c.}{\cX\eta\cbX} - M^{(0)}\Big] \bpl
     \der_\nu \cB_{\l\s} -\eta_{IJ}\cW_\nu^I \der_\l \cW_\s^J
     \bpr \Big\} \, .
\eq}
\bq \cW^1_\mu \mathrel{\mathop{=}\limits^{\rm def}}  V_\mu^d \hspace{2cm}
\cB_{\mu\nu} \mathrel{\mathop{=}\limits^{\rm def}} B_{\mu\nu} +
W_{[\mu}^0 \cW_{\nu ]}
\eq
Note that the gauge transformation of $V_\mu^d$ as given in \eqn{gaugecV}
is consistent with the transformation of $\cW_\mu^1$. Taking into account
the result \eqn{dualgauget} one finds the following gauge transformations
for the antisymmetric tensor theory:
\brr \d_{\rm gauge} \cW_\mu^I &=& \der_\mu \ct^I \nonumber \\
     \d_{\rm gauge} \cB_{\mu\nu}  &=& 2\der_{[ \mu} \check{\Lambda}_{\nu ]}
     + \eta_{IJ}\ct^I \der_{[\mu}\cW_{\nu ]}^J   \, .
\label{delgaugeB}
\err

\subsection{The supersymmetry transformation of the antisymmetric tensor}

The supersymmetry transformation of the antisymmetric tensor field
$\cB_{\mu\nu}$ can be determined from the lagrangian \eqn{vtlagr}. As input
one takes the supersymmetry variations of the background vector multiplet
fields as given in \eqn{vrules} and the variation of $\cW_\mu^1$ in
\eqn{delW1}. Taking into account the redefinitions \eqn{newvar} one finds
that
\bq\begin{array}{lcl}
   \d\phi &=& \be^i\l_i + h.c. \\[2mm]
   \d V^{(z)}_\mu \!\! &=& i\der_\mu \{\be^i\l_i\} + h.c. \\[2mm]
   \d\l_i &=& \bpl\Dslash\phi -i \hat{\Vslash}^{(z)}  \bpr \e_i
   -{\displaystyle \frac{i}{2X^0}} \ve_{ij} \s^{\mu\nu}\e^j \bpl
   \hC'_{\mu\nu} -i \phi  \hF^0_{\mu\nu} + \frac{i}{2} \phi \bar{X}^0
   T^{-\, kl}_{\mu\nu} \ve_{kl}\bpr  \\[2mm]
   && -{\displaystyle \frac{1}{2X^0}} \bpl i Z_{ij} + \phi Y_{ij}^0\bpr
   \e^j -{\displaystyle \frac{1}{X^0}} \bpl \l_i \, \be^j\O^0_j + \O^0_i \,
   \be^j \l_j \bpr \end{array}
\label{dellam}
\eq
with
\brr \hC'_{\mu\nu} &=& C'_{\mu\nu} -  \bpl i\bpsi_{i[ \mu } \g_{\nu ]}
     ( 2X^0 \l_j + \phi \O^0_j ) \ve^{ij} + i \phi X^0 \bpsi_{i\mu}
     \psi_{\nu j} \ve^{ij} +h.c. \bpr \nonumber\\
     \hat{V}_\mu^{(z)} &=& V_\mu^{(z)} - \bpl\ft{i}{2} \bpsi^i_\mu \l_i +
     h.c.\bpr \, .
\err
In order to find the supersymmetry variation of $\cB_{\mu\nu}$ it suffices
to check only those terms in the variation of the lagrangian \eqn{vtlagr}
which would vanish if the Bianchi identity
\bq \e^{\mu\nu\l\s} \der_\mu V_\nu^{(z)} = 0
\eq
would be imposed. There are two different ways to obtain $V_\mu{}^{(z)}$
terms in the variation of the lagrangian. In the first place there is the
$V_\mu{}^{(z)}$ dependence of the lagrangian itself
\brr e^{-1} \Ll_{\rm tensor} &=& e^{-1} \Ll_{\rm tensor}
     \Big|_{V_\mu^{(z)}=0} +\ft{i}{4} e^{-1} \e^{\mu\nu\l\s} V_\mu^{(z)} \,
     \bpl \der_\nu \cB_{\l\s}- \eta_{IJ} \cW_\nu^I \der_\l
     \cW^J_\s \bpr \nonumber\\
     && -\ft{i}{2}  V_\mu^{(z)}\, \cX\eta_I \Dd^\mu \cbX^I
     -\ft{i}{8} V_\mu^{(z)}\, \eta_{IJ} \cbO{}^{iI} \g^\mu \cO^J_i+ h.c.
     \nonumber\\
     && +\ft{i}{4} V_\mu^{(z)}\, \cX\eta_I  \cbO{}^{iI} \g^\nu \g^\mu
     \psi_{\nu i} + \ft{i}{8} e^{-1} \e^{\mu\nu\l\s} V_\mu^{(z)}\,
     \cX\eta\cbX \bpsi^i_\nu \g_\l \psi_{\s i} + h.c. \, ,
\label{depend}
\err
from which the relevant contributions to $\d e^{-1} \Ll_{\rm tensor}$ can
be derived in a straightforward way. Secondly there are the $\l_i$ terms
which vary into $V_\mu{}^{(z)}$ due to $\d \l_i = -i \hat{\Vslash}{}^{(z)}
\e_i + \cdots $. Taking everything together one finds that the
antisymmetric tensor action is invariant under supersymmetry if one defines
\bq \d \cB_{\mu\nu} = - 2 \be_i \s_{\mu\nu} \cX\eta_I \cO^{iI}
    - \eta_{IJ} \cW^I_{[\mu} \bpl \be_i \g_{\nu ]} \cO^J_j
    \ve^{ij} +2 \cX^I \be_i \psi_{\nu ] j} \ve^{ij} \bpr
    - 2 \cX\eta\cbX \be^i\g_{[\mu}\psi_{\nu ]i} +h.c.
\label{checkres}
\eq
The $S$-supersymmetry invariance of the theory is automatically guaranteed
so $\cB_{\mu\nu}$ doesn't transform under $S$-supersymmetry. The covariant
field strength for $\cB_{\mu\nu}$ can be determined from the equations
(\ref{delgaugeB}) and (\ref{checkres}):
\bq \hH_{\nu\l\s } = \der_{[ \nu} \cB_{\l\s ]}
    -  \eta_{IJ} \cW^I_{[ \nu}\der_\l \cW^J_{\s ]}
    - \bpl \cX\eta_I  \cbO^{iI} \s_{[\nu\l} \psi_{\s ]i}
    - \ft{1}{2} \cX\eta\cbX \bpsi_{[\nu}^i\g_\l \psi_{\s ]i} +h.c.\bpr \, ,
\eq
while the dual of $\hH_{\mu\nu\l }$ is defined as
\bq \hH^\mu \mathrel{\mathop{=}\limits^{\rm def}} \ft{i}{2} e^{-1}
    \e^{\mu\nu\l\s} \hH_{\nu\l\s }
\eq
The bosonic part of these field strengths is denoted by dropping the
``hat".

It is important to notice that $\cB_{\mu\nu}$ transforms {\em only} into
background Weyl multiplet and vector multiplet fields, and {\em not} for
instance into the dilatini $\l_i$. One can easily understand
why this is the case. At this particular stage of our computation the 
fields $\phi, V_\mu{}^{(z)}$ and $\l_i$ are the only remnants of the 
original vector multiplet based on the scalar $X^1$. They appear at most 
linearly in the lagrangian $e^{-1} \Ll_{\rm tensor}$, which
immediately follows from the linear $X^1$ dependence of the prepotential
(\ref{prep}). The fields $\phi, V_\mu{}^{(z)}$ and $\l_i$ also transform 
at most linearly into eachother, so $e^{-1} \Ll_{\rm tensor}$ can only 
transform into
\bq \ft{i}{4} e^{-1} \e^{\mu\nu\l\s} V_\mu^{(z)} \, \der_\nu \bpl
    \mbox{background fields} \bpr_{\l\s} \, .
\eq
For the other antisymmetric tensor
theories, namely the $\eta_{11}\neq 0$ or $\eta_{1A}\neq 0$ cases
described in section 4 the situation is quite different. The supersymmetry
variation of the tensor $B_{\mu\nu}$ as it was defined in section 4.2 can
be computed much in the same way as $\d \cB_{\mu\nu}$, and one finds that
\bq \d B_{\mu\nu} =  - 4\bpl 2\eta_{11} \phi + \eta_{1A} {\rm Re} Z^A \bpr
|X^0|^2 \be^i \s_{\mu\nu} \l_i  + h.c. + \cdots \, .
\eq
The latter transformation law automatically coincides with what was found
in \cite{wij}. It then suffices to add an independent auxiliary scalar
$\phi^{(z)}$ to recover the off-shell vector-tensor multiplets of
\cite{wij}.

\subsection{Eliminating the auxiliary field $V_\mu^{(z)}$}

Having determined the $Q$-supersymmetry transformation rule for 
$\cB_{\mu\nu}$ we proceed by eliminating the auxiliary field 
$V_\mu{}^{(z)}$. Before actually doing so we briefly recall what happens 
in the vector-tensor multiplet cases of \cite{wij}. There one can 
eliminate $V_\mu{}^{(z)}$ by imposing its own equation of motion, which 
is of the following type: 
\bq V_\mu^{(z)} \sim H_\mu + \cdots \, .
\label{exp}
\eq
A relation like \eqn{exp} is also what we expect to find in the 
present case because it expresses what duality is all about: 
$V_\mu^{(z)}$ was first introduced as the field strength for the axion 
and it should become equal to the dual of the field strength for 
$\cB_{\mu\nu}$ after the dualisation. However, in the heterotic theory 
a relation like \eqn{exp} cannot be recovered in one go. Given 
\eqn{depend} one readily verifies that the equation of motion enforced by
the auxiliary $V_\mu{}^{(z)}$ reads
\brr \hH_\mu &=& i \cX\eta_I D_\mu \cbX{}^I
     +\ft{i}{4}\eta_{IJ}\cbO{}^{iI}\g_\mu\cO^J_i +h.c. \nonumber\\
     &=& \cX\eta\cbX A_\mu + i \cX\eta_I {\hat \der}_\mu
     \cbX^I +\ft{i}{4}\eta_{IJ}\cbO{}^{iI}\g_\mu\cO^J_i +h.c.  \,  ,
\label{Vzeom1}
\err
which cannot be used to solve for $V_\mu^{(z)}$ itself. Instead the r\^ole 
of \eqn{Vzeom1} is to constrain the auxiliary Weyl
multiplet field $A_\mu$, such that it becomes a dependent expression 
involving the dual field strength $H_\mu$. Of course, the $A_\mu$ 
equation of motion \eqn{Teomgen} must be imposed as well
\brr \phi\, \cX\eta\cbX\,  A_\mu &=& \ft{1}{2} \cX\eta\cbX\, V_\mu^{(z)}
     - \bpl \ft{i}{2} \phi\, \cX\eta_I (\der_\mu - b_\mu) \cbX^I +h.c.
     \bpr \nonumber \\
     && + \mbox{fermions}^2 + F^{(0)}\mbox{terms} + \mbox{hypermultiplet
     terms}\, ,
\label{aeom}
\err
and this relation finally determines the value of $V_\mu{}^{(z)}$ as 
a function of $A_\mu$ such that altogether equation \eqn{exp} is indeed 
fulfilled. The fact that we have to eliminate $V_\mu{}^{(z)}$ and
$A_\mu$ both at the same time is of course not completely innocent. 
During the axion dualisation process we loose (another) part of the 
Weyl multiplet. As we already said, this is different in the $\eta_{11} 
\neq 0$ or $\eta_{1A} \neq 0 $ cases of section 4 where the Weyl multiplet
is not affected by the dualisation. This seems to be a crucial property 
which guarantees the existence of the off-shell vector-tensor multiplets of
\cite{wij}. We also
note the different character of the relations \eqn{Vzeom1} and \eqn{aeom}. 
The former just depends on the tree-level part of the theory, whereas the 
latter receives one-loop and hypermultiplet dependent corrections. Both 
relations are $SO(2,n)$ invariant.

We would like to stress that the duality transformations we
have encountered so far, i.e. the one yielding the stringy vectors on the
one hand and the one yielding the heterotic antisymmetric tensor field on
the other hand, are very similar in nature. In both cases the Weyl
multiplet cannot be preserved during the dualisation process such that the
final theories are far from being realised off-shell. This is in
line with the obervations in \cite{CDFVP} and \cite{wij} that the off-shell
superconformal constructions that are known today are not applicable in
case of the stringy vectors or the heterotic antisymmetric tensor theory. 
Of course it cannot really be excluded that there might exist alternative
yet unexplored off-shell superconformal theories, which would reduce
---after a straightforward elimination (in the sense of not involving
any duality transformations) of some well-chosen auxiliary fields--- to
the on-shell stringy vector or antisymmetric tensor theories we
have described so far. We don't know what such off-shell theories would
have to look like, if they exist at all.

\subsection{Implementing the superconformal gauge choices}

So far we described the various ingredients of the heterotic
antisymmetric tensor theory. According to the equations \eqn{checkres} and
\eqn{delW1} the tensor $\cB_{\mu\nu}$ and the vectors $\cW_\mu^I$ do not
transform under supersymmetry into the scalar $\phi$ and the fermions
$\l_i$ but rather into the background fields $\cX^I$ and $\cO^I_i$. In the
present on-shell situation it is in fact quite artificial to call $\cX^I$
and $\cO^I_i$ ``background" fields because due to the relations \eqn{F1eom}
and \eqn{Vzeom1} the gravitational, vector and antisymmetric tensor
variables interfere with eachother such that one can no longer tell which
multiplet serves as a background for the others. Due to the on-shellness
there is also no more reason to keep on using all the variables $\phi$,
$\cX^I$, $\l_i$ and $\cO_i^I$ as independent degrees of freedom. In order
to reduce the number of matter degrees of freedom then, we go to the
Poincar\'e version of the heterotic antisymmetric tensor theory. The
general strategy for going from a superconformal to a Poincar\'e theory has
been discussed in section 2.2, so we can simply apply the general rules to
the case at hand.

As before the lagrange multipliers $D$ and $\chi_i$ enforce the constraints
\eqn{cons} which can be used to restrict the hypermultiplet variables. When
these constraints are taken into account the Einstein term in the action
reads
\bq  \ft{1}{2} XN\bX\, R = \ft{1}{2} \left\{ \phi \cX\eta\cbX + \ft{i}{2}
     \bpl F_I^{(0)} \bX^I - h.c. \bpr \right\} R
     = \ft{1}{2} \phi_{\rm inv} \cX\eta\cbX\, R
\label{einst}
\eq
Remark that this term does {\it not} depend on the one-loop part of the
theory. It is the relation between the true dilaton $\phi_{\rm inv}$ and
the field $\phi={\rm Re} S$ which is subject to loop-corrections and one
may view the appearance of $F_I^{(0)}$ in \eqn{einst} as (yet another)
indication that $\phi$ should not be identified with the true string loop
counting parameter. In fact one might use the result \eqn{einst} as a
purely supergravity definition of what the true dilaton field is: the field
$\phi_{\rm inv}$ is the only scalar field which 1) makes the Einstein term
(and hence any sensible dilatation gauge choice) loop independent, and 2)
reduces to $\phi$ when loop-effects are neglected. Note that the
requirement of $SO(2,n)$ invariance is not strong enough as a criterion to
select the true dilaton because it would yield both $\phi_{\rm inv}$ and
$\phi_{\rm hol}= {\rm Re}\, S_{\rm hol}$ as possible candidates.

If we would impose the standard Einstein-frame dilatation gauge at this
point (as well as the usual $U(1)$ gauge) we would find that the $\cX^I$
fields would become loop-independent functions of the special coordinates
$\cZ^I$,\footnote{with $\cZ^I = (-i,-\ft{1}{2}i Z^A\eta_{AB}Z^B, Z^A)$}
multiplied by a common dilaton factor. But it is straightforward to see
that this dilaton dependence can be completely removed by choosing an
alternative dilatation gauge, i.e.
\bq \cX\eta\cbX = 1
\label{dil}
\eq
Nicely enough this last gauge choice directly leads to the heterotic 
effective action in the string frame, because one may identify
\bq
\phi_{\rm inv} = \ef{-2}
\label{dilrelat}
\eq
where the expectation value of $\ef{}$ equals the string coupling
constant $g_s$.\footnote{Later on we will see that any one-loop correction
to our theory will be weighted by a relative $\ef{2}$ factor which proves
that the latter identification is indeed the correct one.} The 
$S$-super\-symmetry gauge choice which keeps equation \eqn{dil} invariant 
under $Q$-supersymmetry reads
\bq
\cbX\eta_I \cO_i{}^I = 0 \, .
\label{S}
\eq
As usual the $S$-supersymmetry gauge choice itself is not $Q$-supersymmetric
invariant. As a first step in the computation of the
compensating $S$-transformation we note that
\brr \d\cO_i^{\,I} &=& 2\Dslash \cX^I \e_i + \ve_{ij}\s^{\mu\nu} \e^j
     \bpl \hcF_{\mu\nu}^{-I} - \ft{1}{4} \cbX^I T_{\mu\nu}^{-\, kl}
     \ve_{kl} \bpr  +\cY_{ij}^{\,I}\e^j \nonumber\\
     && +2\cX^{I}\eta_i  - \ft{i}{2}  \Lambda_{\rm U(1)} \,
     \cO_i^{\,I} \, ,
\label{delcO}
\err
where by definition $\cY_{ij}^{\,0} = Y_{ij}^{\,0}$ and $\cY_{ij}^{\,A}
= Y_{ij}^{\,A}$ while $\cY_{ij}^{\,1}$ is fixed by the constraint
\bq \cbX\eta_I \cY_{ij}^{\,I} = \ft{1}{2} \eta_{IJ} \cbO{}^{kI}\cO^{lJ}
    \ve_{ik}\ve_{jl} \, .
\eq
Using this result the compensating $S$-supersymmetry transformation
can be determined to be
\brr \eta_i (\e) &=& -\ft{i}{2} \hHslash \e_i -\ft{1}{2} \ve_{ij}\s^{\mu\nu} \e^j
     \cbX\eta_I \hcF_{\mu\nu}^{-I} - \ft{1}{4} \e^j
     \eta_{IJ} \cbO{}^{kI}\cO^{lJ} \ve_{ik}\ve_{jl} \nonumber\\
     &&+ \ft{1}{4} \g^\mu\e_j \bl \eta_{IJ} \cbO{}^{jI}\g_\mu\cO^J_i
     - \d^j{}_i \eta_{IJ} \cbO{}^{kI}\g_\mu\cO^J_k\br \, .
\label{valeta}
\err
In order to obtain the equations \eqn{delcO} and \eqn{valeta} the dependent
expressions for $T_{\mu\nu\, ij}^+$ \eqn{F1eom} and $A_\mu$ \eqn{Vzeom1}
were freely used. As we already discussed it is only thanks to the fact
that we dualised the $STU$ vector $W^1_\mu$ and the axion $a$ that we could
make the dependent expressions for $T_{\mu\nu\, ij}^+$ and $A_\mu$ ---and
hence also $\eta_i (\e)$--- completely loop and hypermultiplet independent.
Note that $\eta_i (\e)$ depends on the gauge fields, so the $\cF_{\mu\nu}
{}^I$ and $H_\mu$ dependence of various supersymmetry transformation laws
changes by going to the Poincar\' e theory.  
  
Fixing the $U(1)$ gauge leads to the following expressions for the
dependent scalars $\cX^I$, the gauginos $\cO^I_i$ and the compensating
$U(1)$ transformation respectively
\brr \cX^I &=& (\cZ\eta\cbZ)^{\textstyle -\frac{1}{2}} \cZ^I \nonumber\\
     \cO_i^I &=& (\cZ\eta\cbZ)^{\textstyle -\frac{1}{2}}
     \left( \cL_i^I -  \cZ^I \frac{\cbZ\eta_J \cL_i^J}{\cZ\eta\cbZ}
     \right) \nonumber\\
     \Lambda_{\rm U(1)} (\e) &=& \ft{i}{2} \frac{\cbZ\eta_I
     \be^i\cL_i^I}{\cZ\eta\cbZ} + h.c. \, ,
\label{u1res}
\err
where as usual the $\cL^I_i$ are defined as the fermionic partners of the
special coordinates $\cZ^I$.

Now we concentrate on the hypermultiplet variables. Given the
conditions \eqn{cons}, \eqn{dil} and \eqn{S} we have that
\brr \ft{1}{2} A_i{}^\a A^i{}_\b d_\a{}^\b &=& - \ef{-2} \nonumber\\
      A_i{}^\a \zeta_\b d_\a{}^\b &=& \ef{-2} \vartheta_i
\label{cons2}
\err
where the $SO(2,n)$ invariant dilatini $\vartheta_i$ are defined as
\bq \vartheta_i \mathrel{\mathop{=}\limits^{\rm def}}
    -\ft{1}{2}  \ef{2} \l_i
    - \ft{i}{4} \cO_i^J (\cZ\eta\cbZ)^{\textstyle -\frac{1}{2}} \ef{2}
    \left\{ \bar{Z}^I\bpl F_{IJ}^{(0)} - h.c.\bpr
    - \frac{\cbZ\eta_J }{\cZ\eta\cbZ}\bpl \bar{Z}^K F_K^{(0)}(Z)
    - h.c.\bpr \right\} \, .
\label{defvarth}
\eq
The string-frame dilatation gauge choice is directly responsible for the
dilaton-dilatini dependence at the right hand side of  \eqn{cons2}. The
$\varphi$ and $\vartheta_i$ dependence would have been moved from the
hypermultiplet sector to the vector multiplet sector if we would have
imposed the alternative Einstein frame dilatation gauge. In the present
string frame we may split the hypervariables into a dilaton-dilatini part
and the rest:
\brr A_i{}^\a &=& \ef{-} {A^\prime}_i{}^\a \nonumber\\
     \zeta^\a &=& \ef{-} \bpl {\zeta^\prime}^\a - {A^\prime}_i{}^\a
     \vartheta^i \bpr
\err
Fixing the SU(2) gauge then leads to
\brr {A^\prime}_i{}^\a &=& \sqrt{\frac{-2}{C(B)}} \,\d_i{}^s B_s^\a
     \hspace{1cm} s = 1,2 \hspace{1cm} C(B) = B_s{}^\a B^s{}_\b d_\a{}^\b
     \nonumber\\
     {\zeta^\prime}^\a &=& \sqrt{\frac{-2}{C(B)}} \left( \xi^\a -
     B_s{}^\a
     \frac{2 B^s{}_\b\, d^\b{}_\g \xi^\g}{C(B)} \right) \nonumber\\
     \Lambda_i{}^j &=& 2 \be_i \vartheta^j + 4 \bar{\e}_i \d^j{}_s
     B^s{}_\a\, d^\a{}_\b \xi^\b C(B)^{-1} - h.c.;\mbox{traceless}
\err
where $B_s^\a$ and $\xi^\a$ are the physical hypervariables. Special
coordinates on the quaternionic manifold are defined by splitting the index
$\a$ into the values $1,2$ and the rest, and by putting
\bq B_s{}^{\a =1,2} = \d_s{}^{1,2} \qquad\qquad \xi^{1,2} = 0\, .
\eq
In that case
\brr && \sqrt{\frac{C(B)}{-2}} A_i{}^{1,2} = \ef{-} \d_i{}^{1,2}
     \nonumber\\
     && \sqrt{\frac{C(B)}{-2}} \zeta^{1,2} = - \ef{-} \vartheta^{1,2}
     - 2 \ef{-} \d^{1,2}_s \frac{B^s{}_\a\, d^\a{}_\b \xi^\b}{C(B)} \, .
\err
which indicates that after the Poincar\'e reduction to the string frame
the compensating hypermultiplet $(A_i{}^{1,2},\zeta^{1,2})$ describes the
dilaton and dilatini degrees of freedom. This is the $N=2$ analog of the
$N=1$ statement \cite{BS,dBS} that in $N=1$ $d=4$ heterotic string
effective actions the dilaton can be viewed as sitting in the real part of
a compensating chiral + antichiral multiplet.

It remains to integrate out the auxiliary field $\Vv_\mu{}^i_{\, j}$.
When writing the second equation of \eqn{Teomgen} in a manifestly
$SO(2,n)$ invariant way, and taking into account the various
constraints which have been imposed on the vector and
hypermultiplet variables we get that
\brr \Vv_\mu{}_i{}^j
     &=&  -\bpl {A^\prime}_i{}^\a \hat{\der}_\mu {A^\prime}^j{}_\b
     d_\a{}^\b - {A^\prime}^j{}_\b \hat{\der}_\mu {A^\prime}_i{}^\a
     d_\a{}^\b \bpr \nonumber\\
     &&- \ft{1}{2} \bpl \eta_{IJ} -\ef{2}{\rm Im} F_{IJ}^{(0){\rm cov}}
     \bpr \bpl  \cbO{}^I_i\g_\mu \cO{}^{jJ} - \ft{1}{2} \d_i{}^j
     \cbO{}^I_k\g_\mu \cO{}^{kJ} \bpr
\label{valueV}
\err
Remark that together with the $Y_{ij}^I$ equation of motion \eqn{Yeom}
this equation is the only one ---out of all the equations of motion or
gauge choices we have imposed so far on the antisymmetric tensor theory---
which does involve the one-loop part of the theory. In \eqn{valueV} the
following covariantised second derivative matrix appears
\brr F_{IJ}^{(0){\rm cov}} &=& F_{IJ}^{(0)}
     -\eta_{IJ} \frac{F_K^{(0)} \bX^K}{\cX\eta\cbX}
     -2 \frac{ \cX\eta_{(I}^{{\phantom (}} \bpl F_{J)K}^{(0)} -h.c.\bpr
     \bX^K}{\cX\eta\cbX} \nonumber\\
     && + \bX^K \bpl F_{KL}^{(0)} -h.c.\bpr \bX^L \frac{\cX\eta_{(I}
     \cX\eta_{J)}}{(\cX\eta\cbX)^2} + 2 \bpl F_K^{(0)} \bX^K -h.c.
     \bpr \frac{\cX\eta_{(I} \cbX\eta_{J)}}{(\cX\eta\cbX)^2} \, .
\err
This (non-holomorphic!) function $F_{IJ}{}^{(0){\rm cov}}$ is a natural
object in terms of which many one-loop properties of the heterotic
effective action can be expressed\footnote{
The following identities are handy in explicit computations:
\brr &&F_{IJ}^{(0){\rm cov}} \cX^J = F_I^{(0)}
     -\cX\eta_I \frac{F_K^{(0)} \bX^K}{\cX\eta\cbX} \nonumber\\[-1mm]
     &&\cX^I F_{IJ}^{(0){\rm cov}} \cX^J = F^{(0)} \nonumber\\
     &&\cbX^I F_{IJ}^{(0){\rm cov}} \cX^J = 0 \, \nonumber.
\err  }. As can be verified by a straightforward but tedious computation
$F_{IJ}{}^{(0){\rm cov}}$ transforms as a tensor under the $SO(2,n)$
transformations (apart from a non-trivial monodromy transformation):
\brr F_{IJ}^{(0){\rm cov}}
     &\mathrel{\mathop{\longrightarrow}\limits^{\rm SO(2,n)}}&
     \bpl F_{KL}^{(0){\rm cov}}  + \Lambda_{KL} \bpr
     (\Uu^{-1})^K{}_I  (\Uu^{-1})^L{}_J
     - \eta_{IJ} \frac{\cX\Lambda\cbX}{\cX\eta\cbX} \, .
\err
At this point we have finished the construction of the Poincar\'e version
of the antisymmetric version of the heterotic effective action and we may
now present the supersymmetry transformation rules for all the fields, as
well as the (bosonic part) of the action. The supersymmetry transformation
rules read
\brr \d e_\mu\,^a &=& \bar{\e}^i\g^a\psi_{\mu i}+h.c. \nonumber\\
     \d\psi_\mu^i &=& 2 \hat{\der}_\mu\e^i + \Vv_\mu{}^i{}_j\e^j
     + \cX\eta_I \hat{\der}_\mu \cbX{}^I \e^i
     - i \s_{\mu\nu} \hH^\nu \e^i + \ve^{ij} \g^\nu \e_j \cX\eta_I
     \hcF_{\mu\nu}^I + \cdots \nonumber\\
     \d \cB_{\mu\nu} &=& -\bpl 2 \be^i\g_{[\mu}\psi_{\nu ]i}
     +\eta_{IJ} \cW^I_{[\mu} \bl \be_i \g_{\nu ]} \cO^J_j
     +2 \cX^I \be_i \psi_{\nu ] j} \br \ve^{ij}  +h.c. \bpr
     + 2\der_{[ \mu} \check{\Lambda}_{\nu ]}+ 
     \eta_{IJ}\ct^I \der_{[\mu}\cW_{\nu ]}^J \nonumber\\
     \d \cZ^I  &=& \e^i \cL_i^I   \nonumber\\
     \d \cW_\mu^I &=& \bpl \bar{\e}_i\g_\mu \cO_j^{\,I} \ve^{ij}
     +2\cbX{}^I  \bar{\e}^i\psi_\mu^j \ve_{ij}  +h.c. \bpr + \der_\mu
     \ct^I  \nonumber\\
     \d\cO_i^{\,I} &=& 2\bpl \hat{\der}_\mu \cX^I - \cX^I \cbX\eta_J
     \hat{\der}_\mu \cX^J \bpr\e_i + \ve_{ij}\s^{\mu\nu} \e^j
     \bpl \hcF_{\mu\nu}^I - 2 {\rm Re} \, (\cX^I \cbX\eta_J)
     \hcF_{\mu\nu}^J \bpr  +\cY_{ij}^{\,I}\e^j
     + \cdots \nonumber\\
     \d \varphi &=& \e^i \vartheta_i + h.c. \nonumber\\
     \d\vartheta_i &=& \bpl \hat{\dslash} \varphi + \ft{i}{2} \hHslash  
     \bpr \e_i
     + \ft{1}{2}\Vvslash{}_i{}^j\e_j +  {A^\prime}_i{}^\a \hat{\dslash}
     {A^\prime}^j{}_\b  d_\a{}^\b \e_j + \ft{1}{2} \ve_{ij}\s^{\mu\nu}\e^j
     \cbX\eta_I \hcF_{\mu\nu}^I + \cdots \nonumber\\
     \d B_s{}^\a &=& \bpl 2\bar{\xi}^\a\e_i
     +2\rho^{\a\b}\ve_{ij}\bar{\xi}_\b\e^j\bpr \d^i{}_s \nonumber\\
     \d {\zeta^\prime}^\a &=& \bpl \hat{\dslash} {A^\prime}_i{}^\a
     + {A^\prime}_j{}^\a {A^\prime}^j{}_\b d^\b{}_\g \hat{\dslash}
     {A^\prime}_i{}^\g \bpr \e^i +\cdots
\label{del}
\err
Let us discuss a few aspects of the equation \eqn{del}. Almost all
transformation laws are completely saturated at the string tree level. The
only exceptions are $\d\psi_\mu^i, \d\cO_i^{\,I}$ and $\d\vartheta_i$,
because these contain $\Vv_\mu{}^i{}_j$ and $\cY_{ij}^{\,I}$ terms. As we
already said the latter generate an $F^{(0)}$ dependence which is of higher
order in the fermions. The $\cdots$ stand for other higher-order fermion
terms which are not affected by string loop effects. These $\cdots$ terms
are not particularly interesting, and in any case they can be reconstructed
on the base of the results we presented before. It is particularly
interesting to see that the antisymmetric tensor $\cB_{\mu\nu}$ is still
completely decoupled from the dilaton-part of the theory. Rather
then transforming into $\vartheta_i$ the antisymmetric tensor transforms
into the gravitinos $\psi_\mu^i$, so in the string frame $\cB_{\mu\nu}$ has
become part of the gravitational multiplet. The gravitinos themselves
transform back into the antisymmetric tensor, thanks to the compensating
$\eta_i(\e)$ transformation. The $\d\cO_i^{\,I}$ are $H_\mu$ independent
because there is a cancellation between two contributions coming from the
$Q$- and the $S$-supersymmetry sectors respectively. The variation of the
dilatini $\vartheta_i$ is most easily obtained by varying the left hand
side of the second line of  \eqn{cons2}, although it can also be
computed directly from \eqn{dellam} and \eqn{defvarth}. Note that the
complete $\Vv_\mu{}^i{}_j$ and $\eta_i (\e)$ dependence of $\d\zeta^\a$
feeds into $\d\vartheta_i$ and not into $\d{\zeta^\prime}^\a$. As a result
the latter is $H_\mu$, $\cF_{\mu\nu}^I$ and loop-independent as expected.

The antisymmetric tensor lagrangian reads
\brr e^{-1} \Ll_{\rm tensor} &=& \ef{-2} \left\{ \ft{1}{2} R
     +2 \der^\mu \varphi\, \der_\mu\varphi
     + \ft{1}{4} H^\mu H_\mu + G_{I\bar{J}} \der^\mu \cZ^I \der_\mu \cbZ{}^J
     + \Delta_\a{}^\b \der^\mu B_s{}^\a  \der_\mu B^s{}_\b \right\}
     \nonumber\\[1mm]
     && + \ef{-2} \left\{  \ft{1}{8}  {\cal N}_{IJ} \cF_{\mu\nu}^{+I}
     \cF^{\mu\nu +J}  +h.c. \right\}\nonumber\\
     &&- \ft{1}{2} H^\mu \der_\mu \cbZ{}^I
     \frac{ F_{IJ}^{(0) {\rm cov}} \cZ^J} {\cZ\eta\cbZ}
     + \ft{1}{4} H^\mu \der_\mu M^{(0)} + h.c.
\label{theaction}
\err
with
\brr G_{I\bar{J}} &=& \frac{\eta_{IJ}}{ \cZ\eta\cbZ}
     - \frac{\cbZ\eta_I \cZ\eta_J}{ (\cZ\eta\cbZ)^2} + \ft{i}{2}
     \frac{F_{IJ}^{(0) {\rm cov}} - h.c.}{\cZ\eta\cbZ} \, \ef{2}\nonumber\\
     \Delta_\a{}^\b &=& \frac{2}{C(B)} d_\a{}^\b - \frac{4}{C(B)^2}
     (B^r{}_\g d^\g{}_\a) (B_r{}^\d d_\d{}^\b) \nonumber\\
     {\cal N}_{IJ} &=& \bpl \eta_{IJ} - 4 \frac{\cZ\eta_{(I}
     \cbZ\eta_{J)}}{ \cZ\eta\cbZ} \bpr -i
     \bpl \bF_{IJ}^{(0) {\rm cov}} +  M^{(0)} \eta_{IJ} \bpr \,\ef{2}
\err
This lagrangian is manifestly $SO(2,n)$ invariant except for the ${\cal
N}_{IJ}$ term which generates a shift in the $\theta$-angles according to
\eqn{respons}. The dilaton almost completely decouples from the other
fields except from the familiar $\ef{-2}$ prefactor appearing at the string
tree level. The antisymmetric tensor field also largely decouples from the
rest, although it starts to interact with the vector multiplet moduli
$\cZ^I$ at one loop.  Of course these facts were already known from string
theory, but here we see that we can reproduce them on the base of $N=2$
supersymmetry only. We want to draw attention to the fact that the dilaton
kinetic term (with the characteristic +2 prefactor ) is entirely generated
by the compensating hypermultiplet contribution to the $- \Dd_\mu A_i{}^\a
\Dd^\mu A^i{}_\b  d_\a{}^\b$ term in the original superconformal action.
Had we imposed the Einstein frame dilatation gauge, then the dilaton
dynamics (with a -1 prefactor) would have come from the compensating vector
multiplet part of $N_{IJ} \Dd_\mu X^I \Dd^\mu \bX^J$.

In this article we have chosen a string frame dilatation gauge because
then the metric $g_{\mu\nu}$ automatically coincides with the string
metric. Given the string frame results \eqn{del} and \eqn{theaction}
one can of course immediately read off what the Einstein-frame theory
would look like. It suffices to perform an interpolating dilatation and
$S$-supersymmetry transformation on the various fields, which implies
that
\bq \begin{array}{rclrcl} e_\mu\,^a &=& \ef{} e_\mu\,^a{}^{(E)}\hspace{2cm}
    &\psi_\mu^i &=& \eft{} \bpl \psi_\mu^i{}^{(E)} + \g_\mu{}^{(E)}
    \vartheta^i{}^{(E)} \bpr \\
    \cL_i^{\,I} &=& \eft{-} \cL_i^{\,I}{}^{(E)} \hspace{1cm}
    &\vartheta_i &=& \eft{-} \vartheta_i{}^{(E)} \\
    \xi^\a &=& \eft{-} \xi^\a{}^{(E)} \hspace{1cm}
    &\e_i &= & \eft{}  \e_i{}^{(E)}
    \end{array}
\eq
while $\cB_{\mu\nu}, \cZ^I, \cW_\mu^I, \varphi , B_s{}^\a$ are left
invariant. In this way one finds that
\brr \d \cB_{\mu\nu} &=& - \ef{2} \bpl 4 \be^i \s_{\mu\nu} \vartheta_i{}
     - 2 \be^i \g_{[\mu } \psi_{\nu ]i} \bpr^{(E)} 
     - \eta_{IJ} \cW^I_{[\mu} \bpl \be_i\g_{\nu ]} \cO^J_j{}
     +2 \cX^I \be_i \psi_{\nu ] j} \bpr^{(E)} \ve^{ij}
     +h.c. \nonumber\\
     && + 2\der_{[ \mu} \check{\Lambda}_{\nu ]}+ 
     \eta_{IJ}\ct^I \der_{[\mu}\cW_{\nu ]}^J
\err
where
\brr \cX^I{}^{(E)} &=& \ef{} (\cZ\eta\cbZ)^{\textstyle -\frac{1}{2}}
     \cZ^I \nonumber\\
     \cO_i^I{}^{(E)} &=& \ef{} (\cZ\eta\cbZ)^{\textstyle -\frac{1}{2}}
     \left( \cL_i^I{}^{(E)} -  \cZ^I \frac{\cbZ\eta_J \cL_i^J{}^{(E)}}
     {\cZ\eta\cbZ} + 2 \cZ^I \vartheta_i{}^{(E)} \right)
\err
So in the Einstein-frame the antisymmetric tensor field finally {\it
does} transform into the dilatini, and one ultimately recovers what might
be called the (on-shell) heterotic vector-tensor multiplet. Of course one
is really looking at is a spurious dilatino dependence which has to correct
for the fact that one is not working with the true string gravitinos.

\section{Conclusions}

In this paper we presented the antisymmetric tensor version of the 
low-energy effective action for perturbative heterotic strings on 
$K_3\times T^2$. In fact we took a slightly broader point of view and 
showed that the complete set of antisymmetric tensor theories is quite 
restricted, and that in the dual vector multiplet picture they all lead 
to a theory characterised by one of the standard prepotentials given in 
\eqn{genprep}. Contact between the vector multiplet and the
antisymmetric tensor multiplet pictures can be made by performing the 
change of variables \eqn{newvar}, which immediately leads to the 
appearance of (central charge-like) shift symmetries. Out of the general 
class of antisymmetric tensor theories the one relevant for heterotic
strings is then singled out as the only one for which these shift 
symmetries appear to be inessential. In fact one can obtain a completely 
conventional gauge structure for the heterotic effective action by using 
the so-called stringy vector fields instead of the usual $STU$ vectors. As 
such the stringy vectors play a double r\^ole because they were already 
known to make the $SO(2,n)$ invariance manifest at a lagrangian level.

Another remarkable feature of the heterotic antisymmetric tensor theory 
is that it seems to resist an off-shell description. Within the 
superconformal setup which we used, this on-shell character comes about 
when one is forced to eliminate the auxiliary fields $T_{\mu\nu}^{-\, ij}$ 
and $A_\mu$ contained in the Weyl multiplet. On the other hand we have 
seen that the dependent expressions for $T_{\mu\nu}^{-\, ij}$ and $A_\mu$ 
we obtain, are relatively simple and in particular do not depend on
the one-loop part of the heterotic theory. We want to emphasize that the 
loop-independent expressions for the auxiliary fields only arise when one 
uses the stringy vectors $\cW^I_\mu$ and the antisymmetric tensor field
$\cB_{\mu\nu}$ as the fundamental variables. This at the same time explains
why the final Poincar\'e supersymmetric effective action is considerably
simplified by going to the $\cW^I_\mu$ and $\cB_{\mu\nu}$ formulation.

Several times in this article we found examples where $N=2$ supersymmetry 
considerations are sufficient to unique select the most natural variables 
for the heterotic effective action. The heterotic dilaton $\phi_{\rm inv}$
for instance arises as the only $SO(2,n)$ invariant generalisation of 
$\phi = {\rm Re}\, S$ which makes the Einstein term in the superconformal 
theory loop independent. Choosing a string frame formulation for the 
Poincar\'e theory also has a very clear supergravity interpretation: it 
corresponds to imposing a dilatation gauge which makes the stringy scalars
$\cX^I$ dilaton independent. As such one forces the dilaton and dilatini 
to sit in a compensating hypermultiplet, which ---after three bosonic
degrees of freedom have been removed by an $SU(2)$ gauge choice--- indeed
describes one bosonic and 8 fermionic degrees of freedom. When using the 
natural heterotic variables we discussed before one gets a final theory in
which 1) the $SO(2,n)$ symmetry 
is manifestly realised, 2) the supersymmetry transformation laws are loop
independent apart from some higher order fermionic corrections and 3) the
specific couplings of the dilaton and the antisymmetric tensor (as we know 
them from string theory) are easily reproduced.

{\bf Acknowledgement}
I would like to thank P. Claus, B. de Wit, M. Faux, B. Kleijn and
P. Termonia for numerous discussions concerning the vector-tensor
multiplet in its various appearances. I also would like to thank the Belgian 
IISN for financial support.

\end{document}